\documentclass[aps,prd,showpacs,showkeys,nofootinbib,eqsecnum]{revtex4}
\usepackage{amsmath,amssymb}
\usepackage{epstopdf}
\usepackage{graphicx}
\usepackage{psfrag,color}
\usepackage{subfigure}
\psfrag{bi}{\footnotesize$\widetilde B$}
\psfrag{wi}{\footnotesize$\widetilde W$}
\psfrag{hi}{\footnotesize$\widetilde H$}
\newcommand{\be}{\begin{equation}}
\newcommand{\ee}{\end{equation}}
\newcommand{\bea}{\begin{eqnarray}}
\newcommand{\eea}{\end{eqnarray}}

\def\lsim{\mathrel{\rlap{\raise 2.5pt \hbox{$<$}}\lower 2.5pt\hbox{$\sim$}}}
\def\gsim{\mathrel{\rlap{\raise 2.5pt \hbox{$>$}}\lower 2.5pt\hbox{$\sim$}}}

\newcommand{\Neu}[1]{\ensuremath{\widetilde \chi_{#1}^0}}
\newcommand{\Cha}[1]{\ensuremath{\widetilde \chi_{#1}^\pm}}
\newcommand{\Chap}[1]{\ensuremath{\widetilde \chi_{#1}^+}}

\newcommand{\wbar}[1]{\mkern1mu\overline{\mkern-3mu#1\mkern-1mu}\mkern-2mu}

\allowdisplaybreaks 
\psfrag{neu1g}[r][r]{\scriptsize$\Neu 1 \gamma$}
\psfrag{neu1z}[r][r]{\scriptsize$\Neu 1 Z$}
\psfrag{neu1h}[r][r]{\scriptsize$\Neu 1 h$}
\psfrag{stopb}[r][r]{\scriptsize$\tilde t_1 \bar b$}
\psfrag{neu1w}[r][r]{\scriptsize$\Neu 1 W^+$}
\psfrag{stopt}[r][r]{\scriptsize$\tilde t_1 \bar t$}
\psfrag{sto2t}[r][r]{\scriptsize$\tilde t_2 \bar t$}
\psfrag{chiw-}[r][r]{\scriptsize$\Cha 1 W$}
\psfrag{neu2h}[r][r]{\scriptsize$\Neu 2 h$}
\psfrag{chi1w}[r][r]{\scriptsize$\Cha 1 W$}
\psfrag{chi1h}[r][r]{\scriptsize$\Cha 1 h$}
\psfrag{cha1w}[r][r]{\scriptsize$\Cha 1 W$}
\psfrag{cha1h}[r][r]{\scriptsize$\Chap 1 h$}
\psfrag{chi1z}[r][r]{\scriptsize$\Cha 1 Z$}
\psfrag{cha1z}[r][r]{\scriptsize$\Chap 1 Z$}
\psfrag{neu2w}[r][r]{\scriptsize$\Neu 2 W^+$}
\psfrag{stata}[r][r]{\scriptsize$\tilde \tau \tau$}
\psfrag{cha1w}[r][r]{\scriptsize$\Cha 1 W$}
\psfrag{neu2z}[r][r]{\scriptsize$\Neu 2 Z$}
\psfrag{stauv}[r][r]{\scriptsize$\tilde \tau \nu_\tau$}
\psfrag{stauta}[r][r]{\scriptsize$\tilde \tau \tau$}
\psfrag{selle}[r][r]{\scriptsize$\tilde e_L e$}
\psfrag{muo1m}[r][r]{\scriptsize$\tilde \mu_1 \mu$}
\psfrag{stau2t}[r][r]{\scriptsize$\tilde \tau_2 \tau$}
\psfrag{svtl}[r][r]{\scriptsize$\tilde \nu_{\tau_L} \nu_{\tau_L}$}
\psfrag{neu1cc}[r][r]{\scriptsize$\Neu 1 c c$}
\psfrag{n1mumu}[r][r]{\scriptsize$\Neu 1 \mu\mu$}
\psfrag{n1tata}[r][r]{\scriptsize$\Neu 1 \tau\tau$}
\psfrag{neu1ss}[r][r]{\scriptsize$\Neu 1 s s$}
\psfrag{n1bb}[r][r]{\scriptsize$\Neu 1 b b$}
\psfrag{mmN1}{\ $m_{\Neu 1}$}
\psfrag{mmN2}{\ $m_{\Neu 2}$}
\psfrag{mmN3}{\ $m_{\Neu 3}$}
\psfrag{mmN4}{\ $m_{\Neu 4}$}
\psfrag{mn}[c][c]{$m_{\chi^0_i}$ \footnotesize[GeV]}
\psfrag{neu1dbd2}[r][r]{\scriptsize$\Neu 1 \bar d d/\bar s s$}
\psfrag{neu1ubu2}[r][r]{\scriptsize$\Neu 1 \bar u u/\bar c c$}
\psfrag{n1bbb}[r][r]{\scriptsize$\Neu 1 \bar b b$}
\psfrag{n1vebve2}[r][r]{\tiny$\Neu 1 \bar \nu_{e,\mu} \nu_{e,\mu}$}
\psfrag{n1vtaubv}[r][r]{\scriptsize$\Neu 1 \bar \nu_\tau \nu_\tau$}
\psfrag{n1tautau}[r][r]{\scriptsize$\Neu 1 \bar \tau \tau$}
\psfrag{n1ee2}[r][r]{\scriptsize$\Neu 1 \bar e e/\bar \mu \mu$}
\psfrag{sta1t}[r][r]{\scriptsize$\tilde \tau_1 \bar\tau (*)$}
\psfrag{c1ubd4}[r][r]{\tiny$\Cha 1 \bar u d/\bar c s (*)$}
\psfrag{c1w-}[r][r]{\scriptsize$\Cha 1 W^\mp$}
\psfrag{st1tb}[r][r]{\scriptsize$\tilde t_1 \bar t (*)$}
\psfrag{st2tb}[r][r]{\scriptsize$\tilde t_2 \bar t (*)$}
\psfrag{n1udb2}[r][r]{\scriptsize$\Neu 1 u\bar  d/c\bar  s$}
\psfrag{n1eve2}[r][r]{\scriptsize$\Neu 1 \bar e \nu_e/\bar \mu \nu_\mu$}
\psfrag{n1tauv}[r][r]{\scriptsize$\Neu 1 \bar \tau \nu_\tau$}
\psfrag{sta1v}[r][r]{\scriptsize$\bar{\tilde \tau}_1 \nu_\tau$}
\psfrag{smuRm}[r][r]{\scriptsize$\tilde \mu_R \bar\mu (*)$}
\psfrag{selRe}[r][r]{\scriptsize$\tilde e_R \bar e (*)$}
\psfrag{svtvt}[r][r]{\scriptsize$\tilde \nu_\tau \nu_\tau$}
\psfrag{stau2tau2}[r][r]{\scriptsize$\tilde \tau_2 \bar\tau (*)$}
\psfrag{selLe4}[r][r]{\scriptsize$\tilde e_L \bar e /\tilde \mu_L \bar \mu (*)$}
\psfrag{sveve4}[r][r]{\scriptsize$\tilde \nu_{e,\mu} \bar\nu_{e,\mu}$}
\psfrag{st1bb}[r][r]{\scriptsize$\tilde t_1 \bar b$}
\psfrag{st2bb}[r][r]{\scriptsize$\tilde t_2 \bar b$}
\psfrag{sb1t}[r][r]{\scriptsize$\bar{\tilde b}_1  t$}
\psfrag{lep}[r][r]{\hspace{-3mm}\footnotesize  lep}
\psfrag{h114}[r][r]{\scriptsize$m_h\! <\!114$\! GeV\!}
\psfrag{sg}{\footnotesize (a)}
\psfrag{am}{\footnotesize (d)}
\psfrag{i1}{\footnotesize (c)}
\psfrag{i5}{\footnotesize (b)}
\begin{document}
\preprint{HIP-2009-22/TH}\preprint{DO-TH 09/20}
\begin{flushright}
HIP-2009-22/TH\\{DO-TH 09/20}
\end{flushright}
\title{Implications of different supersymmetry breaking patterns for the spectrum and decay of neutralinos and charginos}
\author{Katri Huitu $^1$  
\footnote{\tt{Electronic address: katri.huitu@helsinki.fi}},
Jari Laamanen $^{2,3}$ 
\footnote{\tt{Electronic address: j.laamanen@science.ru.nl}},
P.  N. Pandita $^4$ 
\footnote{\tt{Electronic address: ppandita@nehu.ac.in}}, and 
Paavo Tiitola $^1$ 
\footnote{\tt{Electronic address: paavo.tiitola@helsinki.fi }}}
\affiliation{ $^1$ Department of Physics, and Helsinki Institute of Physics,
P. O. Box 64, FIN-00014 University of Helsinki, Finland}
\affiliation{ $^2$ Theoretical High Energy Physics, 
Radboud University Nijmegen, P.O. Box 9010, 
NL-6500 GL Nijmegen, The Netherlands}
\affiliation{ $^3$ Institut f\"ur Physik,
Technische Universit\"at Dortmund, {D-44221} Dortmund, Germany}
\affiliation{  $^4$ Department of Physics, North Eastern Hill University, Shillong 793 022, India}
\medskip
\begin{abstract}
We consider different patterns of supersymmetry breaking gaugino
masses, and implications of these patterns for the phenomenology of
neutralinos and charginos in models of low energy supersymmetry.  We
outline a general procedure for obtaining approximate values for the
neutralino masses relevant for our analysis, and describe the
constraints on the gaugino mass parameters which follow from the
present experimental limits on the mass of the lightest chargino.
We evaluate an upper bound on the mass of the lightest neutralino
that follows from the structure of the mass matrix in different
models for the gaugino mass parameters.  Using the experimental
lower limit for the chargino mass, we examine the lower bound for
the neutralino masses.  Using a  sum rule for the
squared masses of charginos and neutralinos, we discuss how to
distinguish between different models for the supersymmetry breaking
scenarios.
We discuss in detail the decay modes
of neutralinos and charginos in different models of supersymmetry breaking.  
Our analysis shows that by measuring the masses and decay properties of 
the neutralinos, one can distinguish between different patterns of 
supersymmetry breaking in the gaugino sector.  
We then compare the dark matter characteristics
that arise in different models of supersymmetry breaking.  
\end{abstract}
\pacs{  12.60.Jv, 14.80.Nb, 14.80.Ly}
\keywords{Gaugino masses, supersymmetry breaking}
\maketitle
\section{Introduction} 
Supersymmetry is at present a leading candidate for physics beyond
the standard model~(SM). In supersymmetric models the Higgs sector of
the standard model, which is crucial for its internal consistency, is
technically natural, which makes these models very appealing.  Since
in nature there are no supersymmetric particles with the same mass as
ordinary particles, supersymmetry must be a broken symmetry at low
energies.  The specific mechanism which breaks supersymmetry is
important in determining the masses of superpartners of the SM
particles, and, hence, the experimental signatures of supersymmetry.
It is widely expected that at least some of the superpartners will be
produced at the Large Hadron Collider~(LHC), which has started
its operations.  However, most of the supersymmetric particles that
are likely to be produced at the LHC will not be detected as such,
since they will eventually decay into the lightest supersymmetric
particle~(LSP), which is stable as long as the R-parity~($R_p$) is
conserved.  Thus, the experimental study of supersymmetry involves the
study of cascade decays of the supersymmetric particles to the LSP,
and the subsequent reconstruction of the decay chains.  The LSP in a
large class of supersymmetry breaking models is the lightest
neutralino, which has, therefore, been a subject of intense study for
a long time~\cite{Bartl:1986hp,Bartl:1989ms,Pandita:1994ms,
Pandita:1994vw, Pandita:1994zp, Pandita:1997zt, Choi:2001ww,
Huitu:2003ci}.  A stable lightest neutralino is also an excellent
candidate for dark matter~\cite{Bertone:2004pz}. In view of the
possible production of supersymmetric particles and their subsequent
decay into the lightest neutralino at the LHC, the properties of the
lightest neutralino, and also those of heavier neutralinos and
charginos, which often appear in the cascade decays, are of
considerable importance.  In particular a detailed study of the
lightest neutralino, especially the predictions for its mass, are of
great importance for the supersymmetric phenomenology.


In the minimal supersymmetric extension of the standard model~(MSSM)
at least two Higgs doublets $H_1$ and $H_2$ with hypercharge~$({Y})$
having values $ -1 $ and $+1$, respectively, are required to generate
masses for all the SM fermions and gauge bosons, and to cancel
triangle anomalies.  By the minimal supersymmetric extension of the SM
we here mean the model with minimal particle content and the standard
model gauge group.  The fermionic partners of these Higgs doublets mix
with the fermionic partners of the gauge bosons to produce four
neutralino states $\tilde\chi^0_i, i=1,2,3,4$, and two chargino states
$\tilde\chi^{\pm}_i, i=1,2$. In extended supersymmetric models, there
can be extra neutralino states~\cite{Pandita:1994ms, Pandita:1994vw}.


The masses of the neutralinos and charginos depend, besides other
model parameters, on the soft supersymmetry breaking gaugino masses
corresponding to the $SU(2)_L$ and $U(1)_Y$ gauge groups. At present
there are several models of supersymmetry breaking.  The model of
supersymmetry breaking that has been studied most extensively is the
gravity mediated~\cite{sugra} supersymmetry breaking model.  In this
class of models, supersymmetry is assumed to be broken in a hidden
sector by fields which interact with the SM particles and their
superpartners~(the visible sector) via only the gravitational
interactions. Whereas this mechanism of supersymmetry breaking is
simple and appealing, it suffers from the supersymmetric flavor
problem. On the other hand, in a different class of models~\cite{gmsb},
supersymmetry is broken in a hidden sector and transmitted to the
visible sector via SM gauge interactions of messenger particles.  This
mechanism of supersymmetry breaking, the gauge mediated supersymmetry
breaking, provides an appealing solution to the supersymmetric flavor
problem.  Both, the gravity mediated and gauge mediated supersymmetry
breaking (GMSB) models, have their distinct experimental signatures.
In the GMSB models, the lightest supersymmetric particle is gravitino
instead of a neutralino.


The soft supersymmetry breaking terms in the two breaking mechanisms
described above have contributions originating from the super-Weyl
anomaly via loop effects.  If gravity and gauge mediation of
supersymmetry breaking are somehow suppressed, the anomaly mediated
contributions can dominate, as may happen, {\it e.g.}, in brane
models~\cite{amsb}.  If this happens, then this mechanism of
supersymmetry breaking is referred to as anomaly mediated
supersymmetry breaking (AMSB).  Anomaly mediation is a predictive
framework for supersymmetry breaking in which the breaking of scale
invariance mediates between hidden and visible sectors.


There is another simple pattern of supersymmetry breaking, which is a
hybrid between anomaly mediated supersymmetry breaking and mSUGRA
pattern, that arises from the mirage (or mixed modulus) mediation of
supersymmetry breaking, which has low energy values for masses quite
distinct from either of the two. Mirage mediation is naturally realized
in KKLT-type moduli stabilization~\cite{Kachru:2003aw} and its
generalizations, a well known example being KKLT moduli stabilization
in type IIB string theory~\cite{Kachru:2002he}.


The soft gaugino masses provide a handle for identifying the type of
supersymmetry breaking \cite{Choi:2007ka,Altunkaynak:2009tg, Lowen:2009nr} 
in the gaugino sector.  In \cite{Choi:2007ka} explicit examples of models 
of each of the patterns mentioned above were discussed.  The possibility 
to detect the gaugino mass nonuniversality at the LHC was studied 
in \cite{Altunkaynak:2009tg}.  It is because of
the distinctive patterns of gaugino masses that one is tempted to
believe that neutralinos and charginos are a key in understanding the
supersymmetry breaking mechanism.


In this paper we consider the phenomenology of neutralinos and charginos in
different models for the soft supersymmetry breaking gaugino masses.
This includes the gravity mediated supersymmetry breaking, the anomaly
mediated supersymmetry breaking, and the mirage mediation of
supersymmetry breaking.  In Section II we recall the essential
features of the neutralino mixing and the resulting mass matrix, and
the constraints on the parameters of the mass matrix that follow from
the experimental lower bound on the mass of the lightest chargino. We
then discuss the patterns of the gaugino masses that arise in the different
patterns of supersymmetry breaking in the gaugino sector, and   
compute the neutralino mass spectrum and discuss the distinguishing
features of the spectrum in different patterns of supersymmetry
breaking gaugino masses.  In Section III we discuss a general upper
bound on the mass of the lightest neutralino, and evaluate this bound
for different models for the soft gaugino masses. Here we also discuss
an upper bound on the lightest neutralino mass that follows from
radiative electroweak symmetry breaking, which determines the absolute
value of the Higgs(ino) mass parameter $\mu$.  In Section IV we
discuss and evaluate a sum rule involving the squared masses of the
charginos and neutralinos, and describe how this sum rule can be used
to distinguish between different models of supersymmetry breaking
gaugino masses.  In Section V we study the two-body decays of
charginos and neutralinos in different models, and discuss how the
branching ratios for these decays can be used to distinguish between
the underlying pattern of supersymmetry breaking gaugino masses. In
Section VI we compute and compare the relic abundance of the lightest
neutralinos in different models assuming that the lightest neutralino
is the lightest supersymmetric particle.  We discuss the constraints
imposed on the parameter space by the precise limits on the relic
density obtained by the Wilkinson Microwave Anisotropy Probe~(WMAP)
satellite~\cite{Spergel:2006hy}.  Finally, in Section VII we 
summarize our conclusions.
\section{Chargino and Neutralino Masses and Gaugino Mass Patterns}
In this section we will
describe constraints on the parameters of the neutralino mass matrix
which follow from the present experimental limits on the mass of the
lightest chargino.  We will then discuss the patterns for the gaugino
mass parameters that arise in different supersymmetry breaking
scenarios, and the resulting consequences for the mass of the lightest
neutralino.
For completeness, we have summarized  the neutralino and chargino mass
matrices~\cite{nilles} in the Appendix, and outlined the procedure to obtain 
approximate eigenvalues for the neutralino mass matrix.

\subsection{ Experimental Constraints}

Collider experiments have searched for the supersymmetric partners of
the standard model particles. No supersymmetric partners of the SM
particles have been found in these experiments.  At present only lower
limits on their masses have have been obtained.  In particular, the
search for the lightest chargino state at LEP have yielded lower
limits on its mass \cite{lep-chargino}.
The lower limit depends on the spectrum of the model \cite{Yao:2006px}.
Assuming that $m_0$ is large, the  
limit on the lightest chargino mass  following from 
nonobservation of chargino pair production in $e^+ e^-$
collisions is 
\be M_{\tilde \chi_1^{\pm}} \gsim 103~~{\rm GeV}.  \label{ch-limit}\ee 
The bound depends on the sneutrino mass.  For a sneutrino mass below 200
GeV, the bound becomes weaker, since the production of a chargino pair
becomes more rare due to the negative interference between $\gamma$ or
$Z$ in the $s$-channel and $\tilde\nu$ in the $t$-channel. In the models we
consider, $m_{\tilde\nu}$ is close to $m_0$.  When $m_{\tilde\nu}<200$
GeV, but $m_{\tilde\nu}>m_{\tilde\chi^\pm}$, the limit 
becomes \cite{Yao:2006px}
%
\be M_{\tilde \chi_1^{\pm}} \gsim 85~~{\rm GeV}.  \ee
For the parameters of the chargino mass matrix the limit
(\ref{ch-limit}) implies 
an approximate lower limit~\cite{Abdallah:2003xe,Dreiner:2009ic} 
\be M_2,~~ \mu \gsim 100~~{\rm GeV}.
\label{limits1}
\ee 
The limits Eq.~(\ref{limits1}) on the parameters $M_2$ and $\mu$ are 
found from scanning over the MSSM parameter space and are thus model
independent. 
\subsection{ Gaugino Mass Patterns}
Having constrained the parameters $M_2$ and $\mu$, which enter the
chargino as well as the neutralino mass matrix, we now turn to the
theoretical models for the supersymmetry breaking gaugino mass
parameters $M_1, M_2$, and $M_3$. Theoretically, a simple set of
patterns has emerged for these SUSY breaking parameters, which can be
described as follows.
\subsubsection{Gravity mediated supersymmetry breaking}
The first pattern, which has been the object of extensive studies, is the one
which arises in the gravity mediated supersymmetry breaking 
models~\cite{sugra}. We recall that the soft gaugino masses $M_i$ 
and the gauge couplings $g_i$ satisfy the renormalization group 
equations~(RGE's)~($|M_3| \equiv  M_{\tilde g}$, the gluino mass) 
\bea
16\pi^2\frac{dM_i}{dt} & = & 2 b_i M_i g_i^2, ~~~~~~~~~b_i =
\left(\frac{33}{5}, 1, -3\right), \label{gaugino1}\\
16\pi^2\frac{dg_i}{dt} & = & b_i g_i^3, ~~~~~~~~~~~~~~t=\ln(\mu/\mu_0), 
\label{gauge1}
\eea
at the leading order, where $ i = 1, 2, 3 $ refer to the 
$U(1)_Y, SU(2)_L$ and the $SU(3)$ gauge groups, respectively, and $\mu$
is the renormalization scale with $\mu_0$ as a reference scale. 
Furthermore, $g_1 =\sqrt{\frac{5}{3}}g',\; g_2 = g$, and $g_3$ 
is the $SU(3)_C$ gauge coupling.  In the minimal supersymmetric
standard model with gravity mediated supersymmetry breaking
and with a  universal gaugino mass $m_{1/2}$ at the 
grand unified scale~(GUT), usually referred to as mSUGRA scenario, 
we have the boundary 
conditions~($\alpha_i = g_i^2/4\pi, \, i = 1, 2, 3$)
\bea
M_1 & = &  M_2 = M_3 = m_{1/2}, \label{gauginogut}\\
\alpha_1 & = & \alpha_2 =  \alpha_3 = \alpha_G, \label{gaugegut}
\eea
at the GUT scale $M_G$. The  RGE's (\ref{gaugino1}) and (\ref{gauge1}) 
imply that the soft supersymmetry breaking gaugino masses scale like
gauge couplings:
\bea
\frac{M_1(M_Z)}{\alpha_1(M_Z)} & = & \frac{M_2(M_Z)}{\alpha_2(M_Z)}
= \frac{M_3(M_Z)}{\alpha_3(M_Z)} =  \frac{m_{1/2}} {\alpha_G},
\label{gaugino2}
\eea
which implies that $M_i/g_i^2$ does not run at the one-loop level.
Although in the context of the gravity mediated supersymmetry breaking 
models arbitrary soft gaugino masses are possible, we 
shall here consider the mSUGRA realisation (\ref{gauginogut})
of the gravity mediated supersymmetry breaking scheme.
The relation (\ref{gaugino2}) reduces the three gaugino mass parameters to
one, which we take to be the gluino mass $M_{\tilde g}$. The other
gaugino mass parameters are then determined through 
\bea M_1(M_Z) & =
& \frac{5 \alpha}{3 \alpha_3~\cos^2\theta_W}~M_{\tilde g} ~~\simeq~~
0.14~M_{\tilde g},
\label{m3relation}\\
M_2(M_Z) & = & \frac{\alpha}{\alpha_3~\sin^2\theta_W}~M_{\tilde g}
~~\simeq~~ 0.28~M_{\tilde g} ,
\label{m2relation}
\eea where we have used the value of various couplings at the $Z^0$
mass \bea \alpha^{-1}(M_Z) = 127.9, ~~~~~ \sin^2\theta_W = 0.23, ~~~~~
\alpha_3(M_Z) = 0.12.  \eea 
For the gaugino mass parameters this leads to the ratio
\begin{equation}
M_1 : M_2 : M_3 \simeq 1 : 2 : 7.1.
\label{msugra0}
\end{equation}
This pattern  is typical of any scheme obeying 
Eqs.~(\ref{gaugino1}) and (\ref{gauginogut}).
Note that the masses above are the running masses evaluated at
the electroweak scale, $M_Z$. This  discussion of the gaugino mass 
parameters is valid at tree level. When the radiative corrections 
are included, the ratio for these parameters in  mSUGRA is modified to   
\begin{equation}
M_1 : M_2 : M_3 \simeq 1 : 1.9 : 6.2.
\label{msugra1}
\end{equation}
Using  the  ratio (\ref{msugra1}) and the lower limit (\ref{limits1}), we have
the constraint  
\bea
M_1 & \gsim & 50~ {\rm GeV}, \label{msugra2}
\eea 
in the gravity mediated supersymmetry breaking models.  

It is important to point out here that whereas the mechanism 
of gravity mediated supersymmetry breaking is simple and appealing, 
it suffers from the supersymmetric flavor problem.  On the other 
hand in a different class of models, supersymmetry is broken in 
the hidden sector and transmitted to the visible sector via 
Standard Model gauge interactions of messanger particles.
This mechanism of supersymmetry breaking, the gauge mediated
supersymmetry breaking~\cite{gmsb}, provides an appealing 
solution to supersymmetric flavor problem.
In order to maintain the successful gauge coupling unification of the 
minimal supersymmetric standard model, it is usually assumed that 
the messanger particles form a full GUT multiplet, in which case
the resulting gaugino masses follow the mSUGRA 
pattern~(\ref{msugra0}) as a consequence of the assumption of 
gauge coupling unification at the GUT scale. However, there are 
more general gauge mediated
supersymmetry breaking models~\cite{gengmsb} which allow any hierarchy of 
gaugino masses. However, we will confine here to the minimal
models in which the gaugino masses follow the mSUGRA 
pattern~(\ref{msugra0}).

\subsubsection{Anomaly mediated supersymmetry breaking}
The second pattern of gaugino masses, which is distinct from the
mSUGRA pattern and emerges under theoretical assumptions that are
appealing, arises in the anomaly mediated supersymmetry breaking
models.  Since the soft supersymmetry breaking parameters are
determined by the breaking of the scale invariance, they can be
written in terms of the beta functions and anomalous dimensions in the
form of relations which hold at all energies.  In the minimal
supersymmetric standard model (MSSM), the pure anomaly mediated
contributions to the soft supersymmetry breaking parameters
$M_\lambda$ (gaugino mass), $m_{i}^2$ (soft scalar mass squared), and
$A_y$ (the trilinear supersymmetry breaking coupling, where $y$ refers
to the Yukawa coupling) can be written as
\bea
M_\lambda &=& \frac{\beta_g}{g} m_{3/2},\label{gmass}\\
m_{i}^2 &=& -\frac 14 \left( \frac{\partial \gamma_i}{\partial
g}\beta_g + \frac{\partial \gamma_i}{\partial y}\beta_y\right)
m_{3/2}^2,\label{smass}\\
A_y &=& -\frac{\beta_y}{y} m_{3/2},\label{Amass}
\eea
where $m_{3/2}$ is the gravitino mass, 
$\beta$'s are the relevant $\beta$ functions, and $\gamma$'s are 
the anomalous dimensions of the corresponding chiral superfields.
An immediate consequence of these relations is that supersymmetry
breaking terms are completely insensitive to physics in the
ultraviolet.  The degrees of freedom that are excited at a given energy 
determine the anomalous dimensions and beta functions, thus
completely specifying  the soft supersymmetry breaking parameters at that
energy.  We note that the gaugino masses are proportional to their 
corresponding
gauge group $\beta$ functions with the lightest supersymmetric 
particle being mainly a wino.
However, it turns out that the pure scalar mass-squared anomaly
contribution for sleptons is negative~\cite{RS}.
There are a number of proposals for resolving this  problem of tachyonic
slepton masses \cite{PR,KSS,anomalyfix,JJ,chk},  but some of the 
solutions may spoil the most attractive feature of the anomaly
mediated models, {\it i.e.}, the renormalization group (RG) invariance of
the soft terms and the consequent ultraviolet insensitivity of the
mass spectrum.  A simple phenomenologically  attractive way 
of parametrizing the nonanomaly mediated contributions to the 
slepton masses, so as to
cure their tachyonic spectrum, is to add a common mass parameter $m_0$
to all the squared scalar masses \cite{Gherghetta:1999sw}, 
assuming that such an
addition does not reintroduce the supersymmetric flavor problem. Such an 
addition of a nonanomaly mediated term destroys the attractive
feature of the RG invariance of soft masses.
However, the RG evolution of the resulting model, nevertheless,
inherits some of the simplicity of the pure anomaly mediated
relations.


There are several alternative ways to generate these extra
contributions to the soft squared masses in the anomaly mediated
supersymmetry breaking scheme.  In particular there are models of
supersymmetry breaking mediated through a small extra dimension, where
SM matter multiplets and a supersymmetry breaking hidden sector are
confined to opposite four-dimensional boundaries while gauge
multiplets lie in the bulk.  We note that in this scenario the soft
gaugino mass terms are due to the anomaly mediated supersymmetry
breaking, and, therefore, are governed by~(\ref{gmass}).  On the other
hand, scalar masses get contributions from both anomaly mediation and
a tiny hard breaking of supersymmetry by operators on the hidden
sector boundary.  These operators contribute to scalar masses at one
loop and this contribution is dominant, thereby making all squared
scalar masses positive.  The gaugino spectrum is unaltered, and the
model resembles an anomaly mediated supersymmetry breaking model with
nonuniversal scalar masses~\cite{KK}.

Using Eq.~(\ref{gmass}), we then have the following pattern for the
ratio of the gaugino masses at tree level: 
\bea M_1 : M_2 : M_3 &\simeq & 3.3 : 1 :9, 
\eea 
which, after radiative corrections (assuming $m_{3/2}=40$ TeV)
are included, becomes
\begin{equation}
M_1 : M_2 : M_3 \simeq 2.8 : 1 : 7.1, 
\label{anomaly1}
\end{equation}
in the minimal supersymmetric standard model with anomaly mediated
supersymmetry breaking.  Schemes in which this pattern is realized
require a strict separation of hidden sector that breaks SUSY from the
visible sector of the MSSM. This implies a strong sequestering, and
requires that all supersymmetry breaking fields are sequestered from
the visible sector.  Nevertheless, it may be achieved in certain class
of theories with extra dimensions or a conformal field theory sector.

Using (\ref{limits1}),  and the anomaly pattern of the gaugino masses
(\ref{anomaly1}), we have
\bea
M_1 & \gsim & 280~ {\rm GeV}. 
\eea
This is to be contrasted with the corresponding result (\ref{msugra2})
for the gravity mediated supersymmetry breaking. 
\subsubsection{Mirage mediated supersymmetry breaking}
There is a third simple gaugino mass pattern that arises from the
mirage (or mixed modulus) mediation supersymmetry breaking, which is a
hybrid between anomaly mediated supersymmetry breaking and mSUGRA
pattern, and has low energy values for masses quite distinct from
either of the two. Mirage mediation is naturally realized in KKLT-type
moduli stabilization and its generalizations, a well known example
being KKLT moduli stabilization in type IIB string
theory~\cite{Kachru:2002he}.  Phenomenology and cosmology of mirage
mediation have been studied in \cite{Choi:2006im,
  Choi:2005uz,endo05,falkowski05,baer06,baer,yama, kitano-lhc,
  kawagoe}. Signatures of the scenario at LHC and the spectrum of
neutralino mass in particular have been studied in~\cite{Choi:2007ka, cho3}.
The boundary conditions for the soft supersymmetry breaking terms
that produce the mirage mediation scheme can be written as
\cite{choi1}
\begin{eqnarray}
  M_a&=& M_0 \Big[\,1+\frac{\ln({\wbar M}_{Pl}/m_{3/2})}{16\pi^2} b_a
  g_a^2\alpha\,\Big],\nonumber \\
  A_{ijk}&=&M_0\Big[\,(a_i+a_j+a_k)
  -\frac{\ln({\wbar M}_{Pl}/m_{3/2})}{16\pi^2}(\gamma_i+\gamma_j+\gamma_k)\alpha\,\Big],
  \nonumber \\
  m_i^2&=&M_0^2\Big[\,c_i-\,\frac{\ln({\wbar M}_{Pl}/m_{3/2})}{16\pi^2}
  \theta_i\alpha-\left(\frac{\ln({\wbar M}_{Pl}/m_{3/2})}{16\pi^2}\right)^2\dot{\gamma}_i\alpha^2\,\Big],
  \label{eq:bc1}
\end{eqnarray}
where $M_0\sim 1$ TeV is a mass parameter characterizing the moduli
mediation, ${\wbar M}_{Pl}$ is the reduced Planck mass, $g_a$
are the gauge couplings and $b_{\alpha}$ the corresponding one-loop 
beta function coefficients,  $\gamma_{i}$ are the anomalous dimensions, and
$\alpha={m_{3/2}}/[{M_0\ln({\wbar M}_{Pl}/m_{3/2})}]={\cal O}(1)$ is a
parameter representing the ratio of anomaly mediation to moduli
mediation, and 
\begin{eqnarray}
  \theta_i & = & 4\sum_a g^2_a C^a_2(\phi_i)-\sum_{jk}|y_{ijk}|^2(a_i+a_j+
a_k),
\\
\dot {\gamma}_i & = & 8 \pi^2 \frac{d \gamma_i}{d \log\mu},
\end{eqnarray}
where  $C_2^a(\phi_i)$ is the quadratic Casimir operator for the gauge group
with gauge coupling $g_a$, and has a value $(N^2 -1)/(2N)$
for the $SU(N)$ gauge group.
  
Thus the generic mirage mediation is parametrized by
\begin{equation} 
  M_0,\,\, \alpha,\,\, a_i,\,\, c_i=1-n_i,\,\,
  \tan\beta.
\end{equation} 
The parameter values $c_{i}=a_{i}=1$ and $\alpha=1$ correspond to the
minimal KKLT compactification of type IIB theory with modular weight
$n_{i}=0$, but other parameter values are also possible for different
scenarios, for example choice of $\alpha=2$ with $a_{H_U}=c_{H_U}=0$
and $a_{U_3}+a_{Q_3}=c_{U_3}+c_{Q_3}=0$ can possibly alleviate the fine
tuning problem for the electroweak symmetry breaking \cite{tevmirage}.
In our studies we have used the values $c_{i}=a_{i}=1$.
At low energies, the gaugino masses in mirage mediation can be written
as
\begin{equation}
  \frac{M_a(\mu)}{g_a^2(\mu)}\,=\,\left(1+\frac{\ln({\wbar M}_{Pl}/m_{3/2})}{16\pi^2}
    g_{GUT}^2b_a\alpha\right)\frac{M_{0}}{g_{GUT}^2}.
\end{equation}
This leads to a unification of the soft gaugino masses at the mirage messenger 
scale \cite{mirage2} 
\bea
M_{\rm mir}=M_{GUT}\left(\frac{m_{3/2}}{{\wbar M}_{Pl}}\right)^{\alpha/2},
\eea 
which is lower than GUT scale for positive values of $\alpha$. 
For $g_{GUT}^2\simeq 1/2$ the resulting low energy values
yield the mirage mass pattern 
\begin{equation}
M_1 : M_2 : M_3 \simeq (1 + 0.66 \alpha): (2 + 0.2 \alpha): (6 - 1.8 \alpha).
\end{equation}
When the radiative corrections are included for the supersymmetry
breaking gaugino masses, we obtain
\bea
M_1 : M_2 : M_3 \simeq 1 : 1.5 : 2.1 ~~~~{\rm for}~ \alpha = 1, \\
M_1 : M_2 : M_3 \simeq 1 : 1.2 : 0.92 ~~~~{\rm for}~ \alpha = 2.
\label{mirage1}
\eea
where we have used the value $M_0=1$ TeV. Thus, for the mirage mediation, 
we find
\bea
& M_1 \gsim & 67 \; {\rm GeV} ~~~~{\rm for}~ \alpha = 1, \\
& M_1 \gsim & 83 \; {\rm GeV} ~~~~{\rm for}~ \alpha = 2.
\eea


\subsubsection{Comparison of the patterns of neutralino and chargino masses}
In Table~\ref{neutmasstable} we show the lightest neutralino and chargino
masses, which satisfy the experimental limit for the mass of the
lightest chargino~\cite{Yao:2006px} for a particular parameter
point.  These masses are calculated
using SOFTSUSY(v.3.0.13)~\cite{Allanach:2001kg}.  The absolute value
of the Higgsino mixing parameter is determined by the condition of
radiative electroweak symmetry breaking (REWSB), and thus depends on
the soft scalar mass parameter $m_0$.  The parameter $m_0$ also
enters the radiative corrections through the scalar masses.  The
masses of the neutralinos in Table~\ref{neutmasstable} have been
calculated for $m_0=200$ GeV and $m_0=1$ TeV for both the mSUGRA and
AMSB models,  and $c_{i} = a_{i} = 1$ for the mirage mediation. 
The sign of the $\mu$ parameter was chosen
positive. Changing it to the negative value can reduce the neutralino
masses by a few GeV's, but this may lead to conflict with the $b\to s \gamma $
constraint. In mSUGRA the trilinear $A$-parameter was set to zero. Changing that
to nonzero values also may decrease the lightest neutralino masses by a
few GeV's.  These masses demonstrate the effect of the sfermion 
spectrum on the neutralino and chargino masses for the models that 
we have studied in this paper.  For anomaly mediated supersymmetry breaking,
we see the familiar result that the lightest neutralino
is closely degenerate with the lightest chargino.
The neutralino spectrum in AMSB models is typically  heavier than in 
the case of mSUGRA.  
In the case of mirage mediation, 
the mass difference of the lightest and heaviest neutralino masses is
smaller as compared to this mass difference in the other models.
In addition, especially in the $\alpha =2$ model, the $\mu$-parameter is
smaller as compared to its value in the other models, leading to larger 
mixing of the gaugino and Higgsino components.
The much heavier spectrum in the $\alpha=2$ mirage mediation model is
due to the tachyonic stops in the spectrum for the lighter particles.
\begin{table}
\begin{tabular}{|l|c|c|c|c|}\hline
Parameters & mSUGRA & AMSB & Mirage $\alpha=1$ & Mirage $\alpha=2$ \\
\hline
$\tan\beta = 5$, $m_0=200$ GeV &
(58,105,250,277)& 
(85,245,505,518)& 
&\\
&(103,278)&
(85,518)&
&\\
$\tan\beta = 20$, $m_0=200$ GeV &
(58,104,229,253) & 
(85,237,474,482) & 
&\\
&(103,253)& 
(85,484)&
 & \\
$\tan\beta = 5$, $m_0=1$ TeV &
(55,103,346,363) &
(102,286,629,638)&
(72,85,165,176)&
(163,186,473,489)\\
&(103, 365) & 
(103,640)&
(85,184)&
(174,479)\\
$\tan\beta = 20$, $m_0=1$ TeV &
(58,104,211,240) & 
(103,286,534,541)&
(72,94,173,197) &
(140,161,549,566)\\
& (103,242) &
(103,545)&
(85,202)&
(150,553)\\
\hline
\end{tabular}
\caption{The lower limits on the masses of the four neutralino states
and two chargino states in each model in the form
($m_{\chi_1^0}, m_{\chi_2^0}, m_{\chi_3^0}, m_{\chi_4^0}$) [GeV]
followed by ($m_{\chi_1^\pm}, m_{\chi_2^\pm}$) [GeV],
with the given set of parameters, following from the experimental lower
bound on the mass of the lightest chargino. For the mirage mediation 
model with $\alpha =2$ the limit is not from the chargino mass bound, 
but from the requirement of the nontachyonic spectrum.
\label{neutmasstable}}
\end{table}

\begin{figure}
\psfrag{m2}[c][c]{$M_2(EW)$ \footnotesize[GeV]}
   \centering
\psfrag{mmN1}{\ $m_{\Neu 1}$\hspace{-20mm}(a)}
   \includegraphics{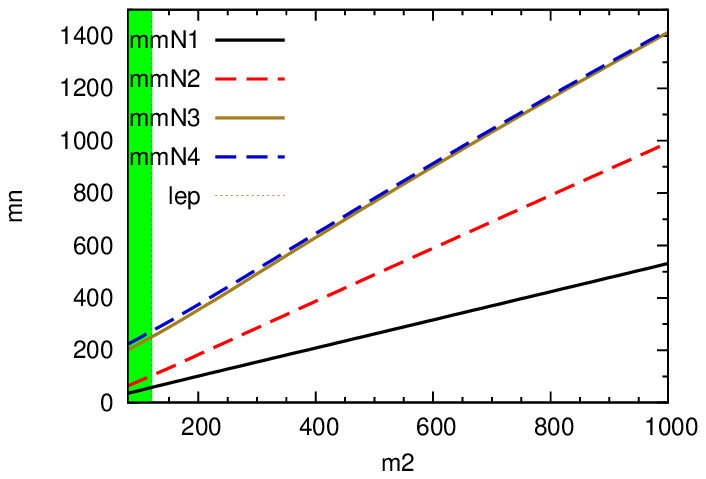}
\psfrag{mmN1}{\ $m_{\Neu 1}$\hspace{-20mm}(b)}
   \includegraphics{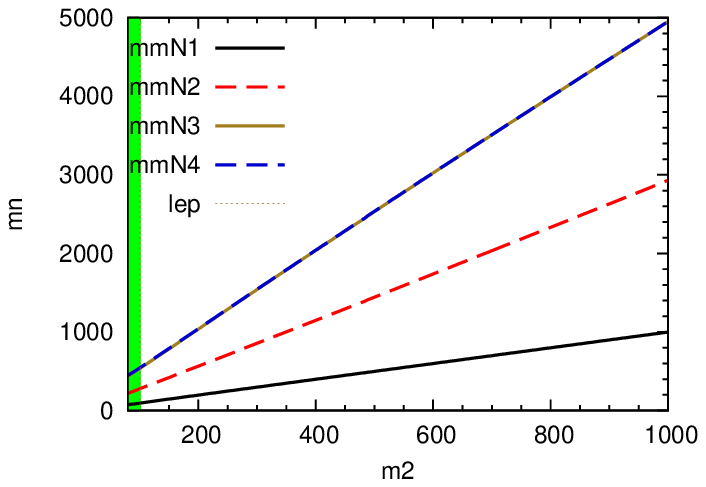}\\
\psfrag{mmN1}{\ $m_{\Neu 1}$\hspace{-20mm}(c)}
   \includegraphics{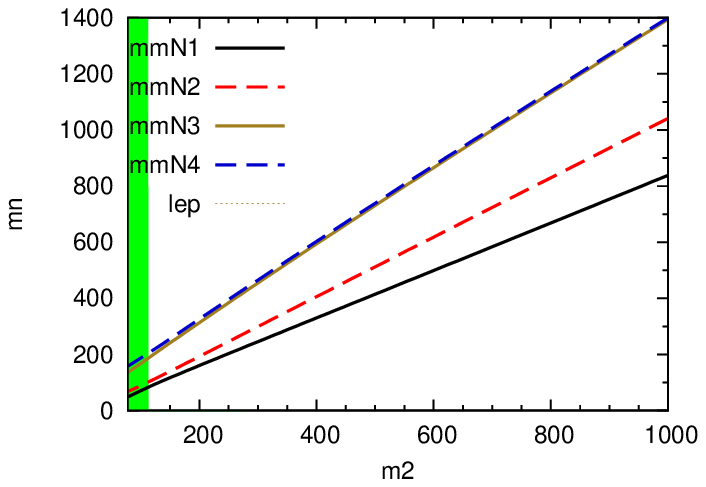}
\psfrag{mmN1}{\ $m_{\Neu 1}$\hspace{-20mm}(d)}
  \includegraphics{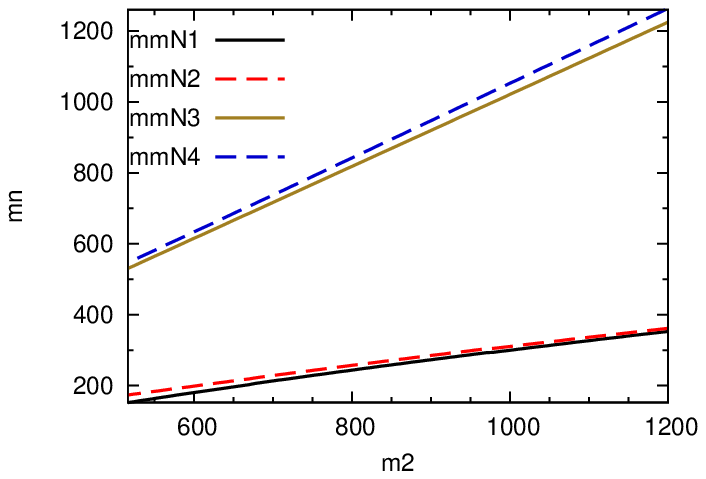}
   \caption{Masses of the neutralinos in the 
     (a) mSUGRA, 
     (b) AMSB, and for
     mirage mediation models  with (c) $\alpha=1$ and 
     (d) $\alpha=2$.  
     Here $\tan\beta = 10,\ \mathrm{sgn} ( \mu ) = +1,\
     A_0=0,\ m_0=1$ TeV (mSUGRA) and 1.5 TeV
     (AMSB). \textsf{lep}-denoted shading indicates the violation of
     the LEP sparticle mass limits.}
   \label{fig:massneutralino}
\end{figure}

The masses of the neutralinos are plotted in
Fig.~\ref{fig:massneutralino} for the mSUGRA, AMSB and the mirage 
mediation scenarios, respectively. 
In mSUGRA the lightest neutralino is more than 80\% bino, while
the second lightest one is more than 70\% wino.
In AMSB, $\chi_1^0$ is almost 100\% wino and $\chi_2^0$ bino.
In the mirage mediation pattern with $\alpha=1$, the compositions of the
two lightest neutralinos are more evenly divided between bino and wino,
but the lightest one is dominantly bino and the second lightest one
wino. For small $M_2$, also the Higgsino component is nonnegligible in
both.  For $\alpha=2$, the $\mu$-parameter becomes relatively small, and both
$\chi_1^0$ and $\chi_2^0$ are more than 90\% Higgsinos.
However, for the small values of $M_2$, the LSP is not a neutralino
as we will discuss later in Sec. VI (see Fig.~\ref{fig:ncMirage}b).


The ratio of the mass parameters, $|M_3|/|M_2|$ is known to
differ drastically in different models.  When the radiative
corrections are taken into account, the ratios for each pattern
discussed above is calculated to be
\bea
\left.\frac{|M_3|}{|M_2|}\right|_{\rm mSUGRA}
=3.3,\; \left.\frac{|M_3|}{|M_2|}\right|_{\rm AMSB}
=7.1,\; \left.\frac{|M_3|}{|M_2|}\right|_{\alpha=1}=1.4,\;
\left.\frac{|M_3|}{|M_2|}\right|_{\alpha=2}=0.77.
\eea
If one applies the ratio to the masses of the particles, one finds
\bea
\left.\frac{|m_{\tilde g}|}{|m_{\chi_1^\pm}|}\right|_{\rm mSUGRA}
= 3.8-4.3,\; \left.\frac{|m_{\tilde g}|}{|m_{\chi_1^\pm}|}\right|_{\rm AMSB}
= 7.3-7.4,\; \left.\frac{|m_{\tilde g}|}{|m_{\chi_1^\pm}|}\right|_{\alpha=1}
= 1.7,\;
\left.\frac{|m_{\tilde g}|}{|m_{\chi_1^\pm}|}\right|_{\alpha=2}=0.9,
\eea
with other parameter values as given  in Table 1.
Thus, very large mass ratio of gluino and chargino hints to an AMSB
type breaking, while small value hints towards mirage type 
of supersymmetry breaking.

\section{The general upper bound on the mass of the lightest neutralino}
In this Section we shall consider a general upper
bound on the mass of the  lightest neutralino that follows from the
structure of the neutralino mass matrix. Since some of the neutralino masses
resulting from diagonalization of the mass matrix~(\ref{neutmatrix})  
can be negative, we
shall for our purposes consider the squared mass matrix $ \hat {\mathcal
M}^{\dagger}\hat {\mathcal M}$. This squared mass matrix can be written as
\bea 
{\mathcal M^{\dagger}_0}  {\mathcal M_0} 
= \left(
\begin{array}{cccc} M_1^2 + M_Z^2  s^2_w  & - M_Z^2 c_w s_w 
& -M_Z  s_w (M_1 c_{\beta} + \mu  s_{\beta}) 
& M_Z s_w(M_1 s_{\beta} + \mu  c_{\beta}) \\
- M_Z^2  c_w s_w & M_2^2 + M_Z^2 c^2_w  & M_Z c_w(M_2 c_{\beta} +\mu s_{\beta})
& -M_Z c_w(M_2 s_{\beta} +\mu c_{\beta})\\
-M_Z  s_w (M_1 c_{\beta} + \mu  s_{\beta}) 
&  M_Z c_w(M_2 c_{\beta} +\mu s_{\beta}) & M_Z^2 c^2_{\beta} + \mu^2
&  M_Z^2 c_{\beta} s_{\beta} \\
M_Z s_w(M_1 s_{\beta} + \mu  c_{\beta}) 
& -M_Z c_w(M_2 s_{\beta} +\mu c_{\beta})  & -M_Z^2 c_{\beta}  s_{\beta} 
&   M_Z^2 s^2_{\beta} + \mu^2
\\
\end{array} \right), \label{neutsquaredmatrix}
\eea 
where $c_W = \cos\theta_W, s_W = \sin\theta_W, c_\beta = \cos\beta$
and  $s_\beta = \sin\beta,$ respectively.
An upper bound on the squared mass of
the lightest neutralino $\chi^0_1$ can be obtained by using the fact
that the smallest eigenvalue of $  {\mathcal M}^{\dagger}_0
{\mathcal M}_0$ is smaller than the smallest eigenvalue of its upper
left $2 \times 2$ sub-matrix
\bigskip
\begin{equation}
\left(
\begin{array}{lr}
M_1^2 + M_Z^2\sin^2\theta_W & -M_Z^2\sin\theta_W \cos\theta_W\\
&\\
-M_Z^2\sin\theta_W \cos\theta_W & M_2^2 + M_Z^2\cos^2\theta_W
\end{array}
\right),
\label{submatrix1}
\end{equation}
\bigskip
\noindent
thereby resulting in the tree-level upper bound~\cite{Pandita:1994zp}
\bea
M_{\tilde \chi^0_1}^2  \le  \frac 12 \left(M_1^2 + M_2^2 +M_Z^2
- \sqrt{(M_1^2 - M_2^2)^2 +M_Z^4 -2 (M_1^2 - M_2^2)M_Z^2\cos
2\theta_W }\right).
\label{bound1}
\eea
We emphasize that the upper bound (\ref{bound1}) is independent of the
supersymmetry conserving parameter $\mu$ and also independent of
$\tan\beta$, but depends on the supersymmetry breaking gaugino mass
parameters $M_1$ and $M_2$. Despite this dependence on the unknown
supersymmetry breaking parameters Eq.~(\ref{bound1}) leads to a useful 
bound on $M_{\chi_1^0}$.
An alternative bound on the mass of the lightest neutralino can be
obtained by considering the bottom-right $ 2 \times 2$ 
sub-matrix
\begin{equation}
\left(
\begin{array}{lr}
M_Z^2 \cos^2\beta + \mu^2  & -M_Z^2 \cos\beta \sin\beta \\
&\\
-M_Z^2 \cos\beta \sin\beta & M_Z^2 \sin^2\beta + \mu^2
\end{array}
\right),
\label{submatrix2}
\end{equation}
leading to an upper bound
\begin{equation}
M_{\chi^0_1}^2 \le |\mu|^2.
\label{bound2}
\end{equation}
\begin{figure}
\psfrag{mch}{$m_{\tilde\chi^\pm_1}$ [GeV]}
\psfrag{mn}{$m_{\Neu 1}$ [GeV]}
\psfrag{mSUGRA}{mSUGRA}
\psfrag{AMSB}{AMSB}
\psfrag{Mirage alpha1}{Mirage mediation, $\alpha=1$}
\psfrag{Mirage alpha2}{Mirage mediation, $\alpha=2$}
\mbox{
       \subfigure[]{\includegraphics[height=6cm]
	{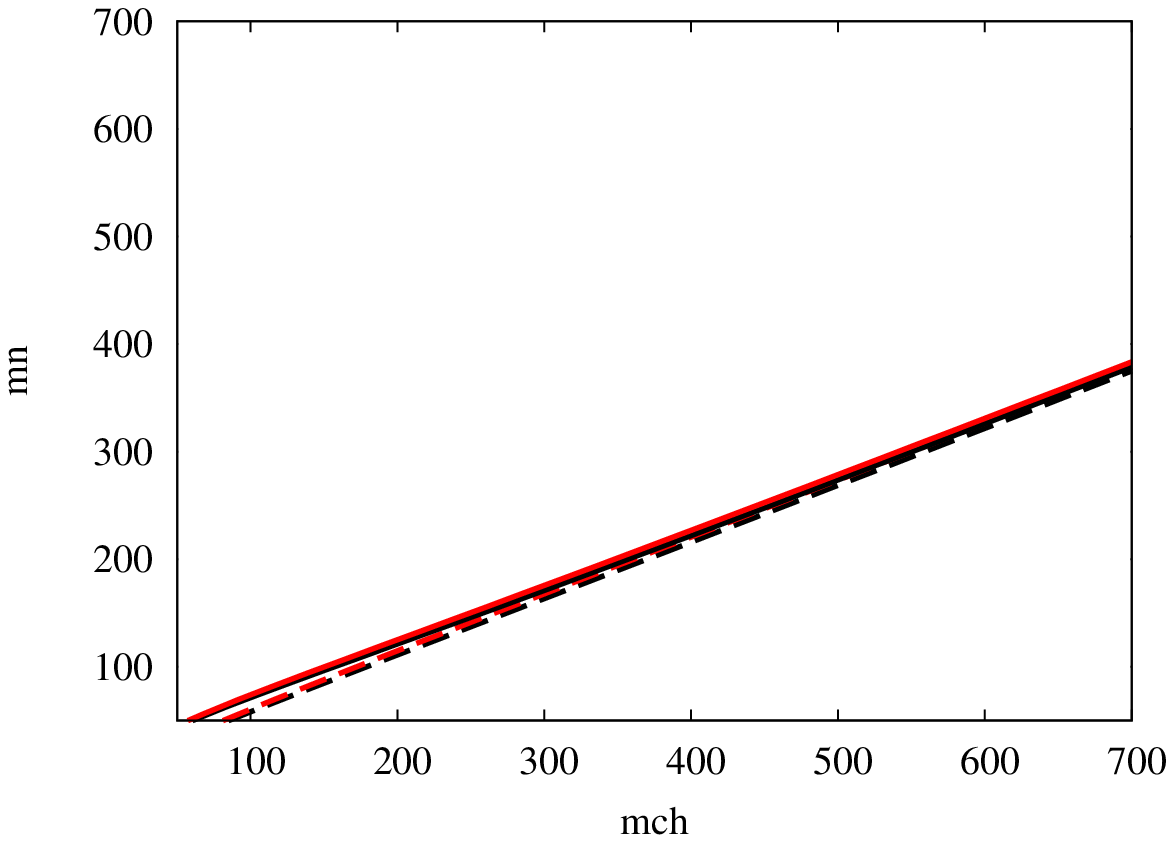}}
        \quad
       \subfigure[]{\includegraphics[height=6cm]
	{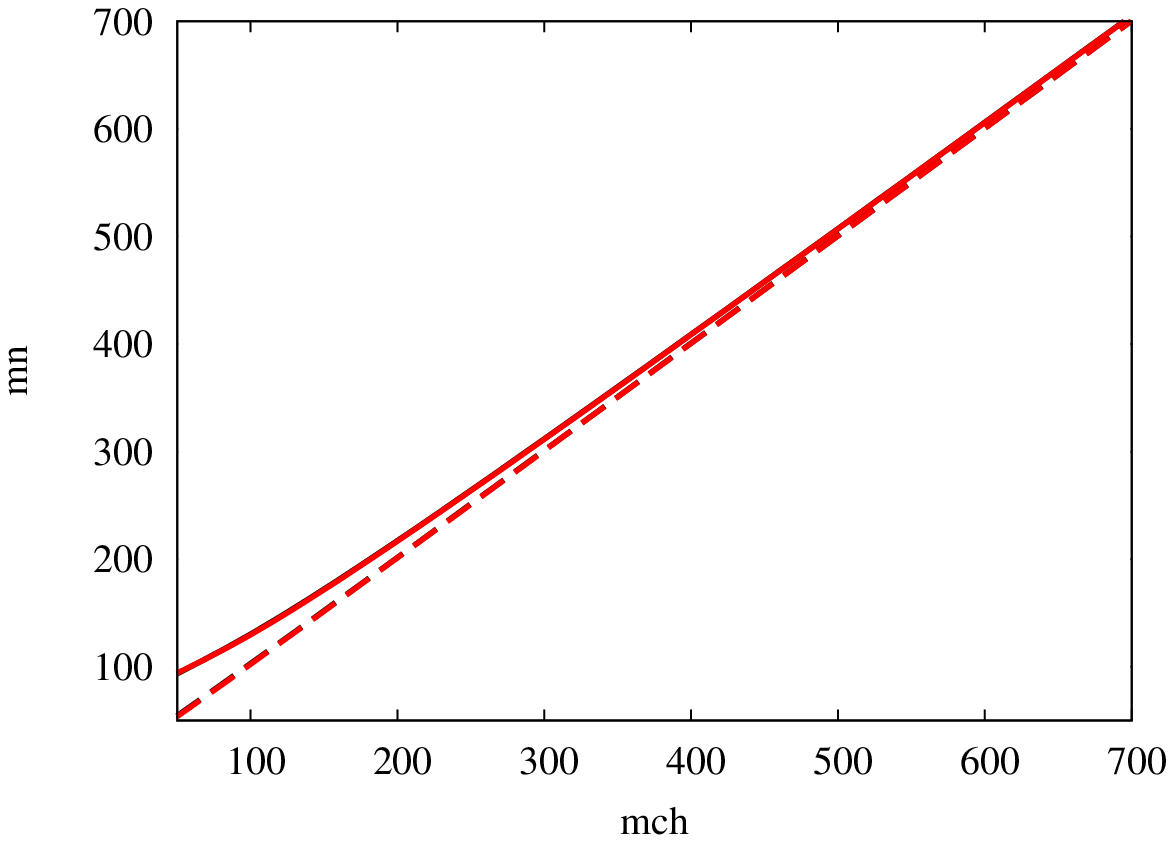}}
 }
\mbox{
       \subfigure[]{\includegraphics[height=6cm]
	{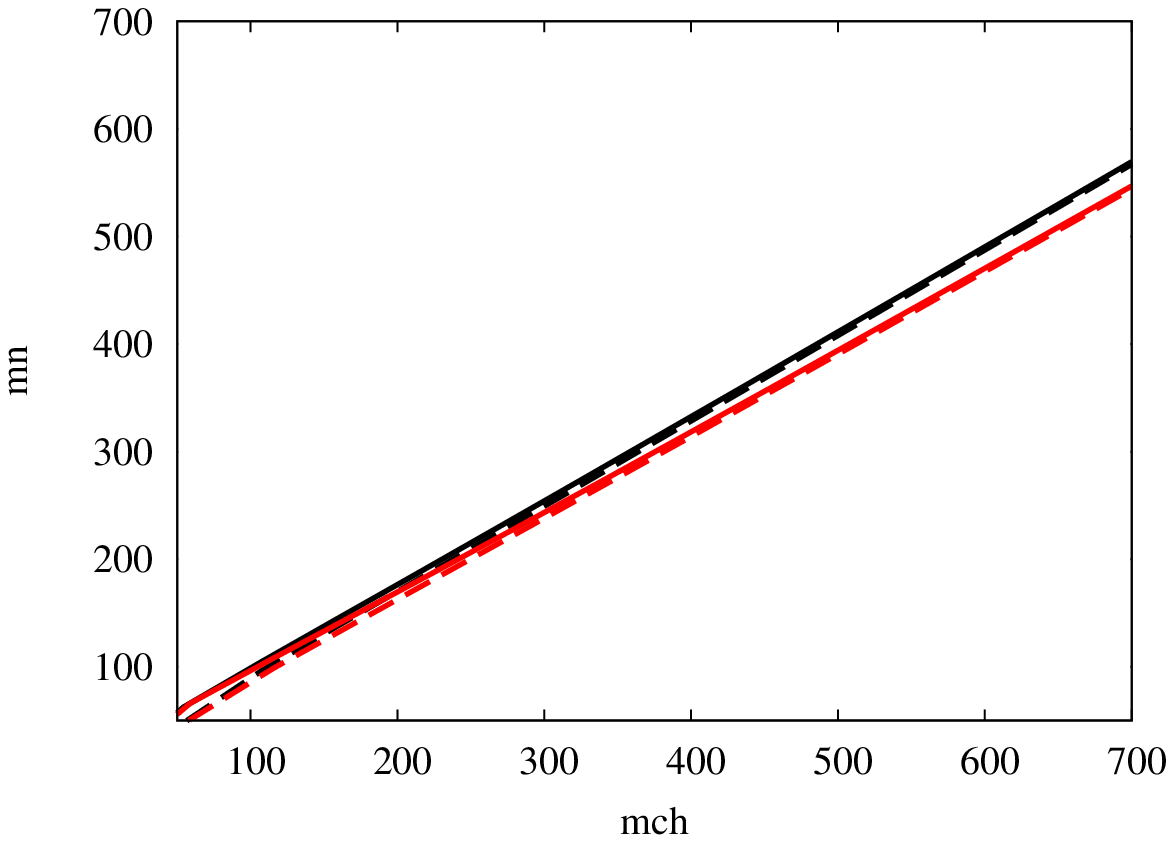}}
        \quad
       \subfigure[]{\includegraphics[height=6cm]
	{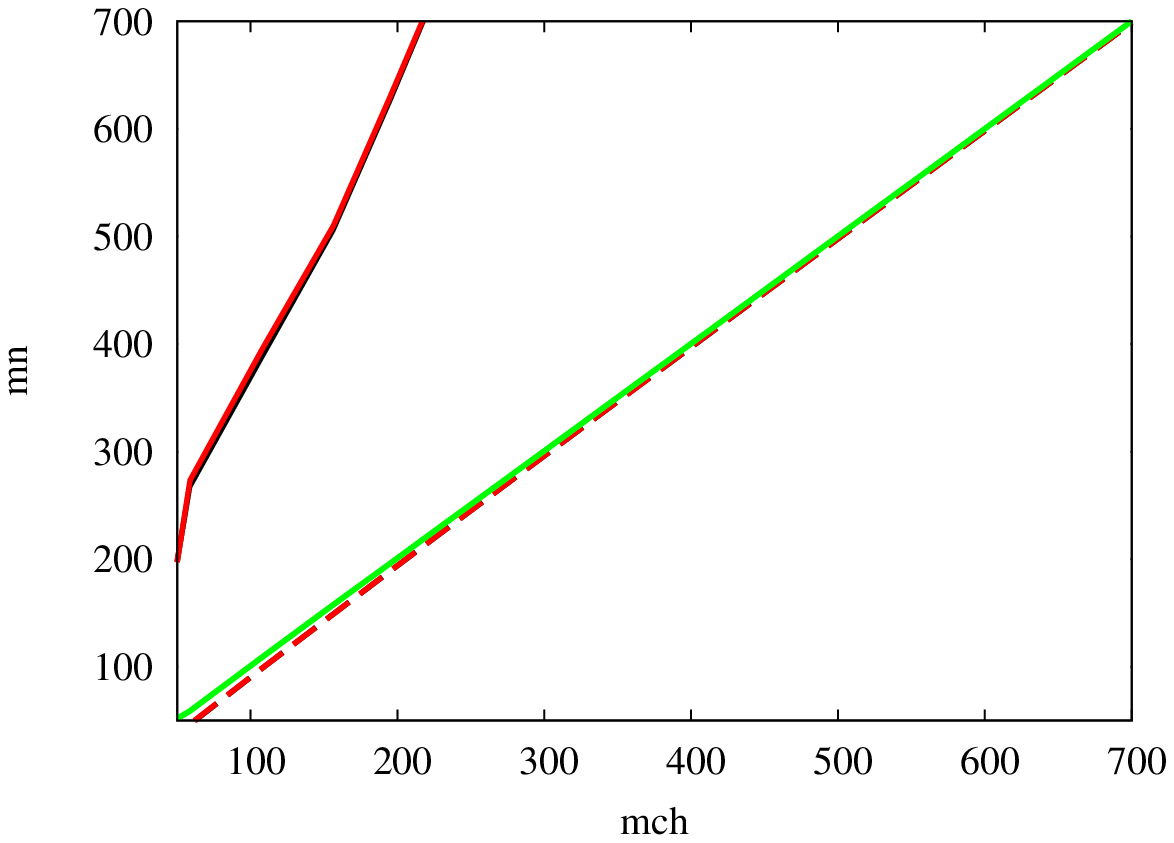}}
 }
 \caption{Upper limit (solid line) and the mass
(dashed line) for the lightest neutralino mass for 
(a) mSUGRA with $m_0=1$ TeV and $A=0$, (b) AMSB with $m_0=1$ TeV, and
mirage mediation with $a=c=1$ and with (c) $\alpha=1$, and (d) 
with $\alpha=2$, calculated at tree-level (red/gray)
and with radiative corrections added (black) as a function of the
mass of the lighter chargino.  In all the plots $\tan\beta =10$
and ${\rm sign}(\mu)=+1$.}
 \label{fig:radcorr}
\end{figure}
\noindent
This bound, when supplemented by  the electroweak symmetry breaking 
condition
\bea
\frac{1}{2} M_Z^2 & = & \frac{(m_1^2 - m_2^2 \tan^2\beta)}
{(\tan^2\beta - 1)} -  |\mu|^2,
\label{ewsymmetry}
\eea
leads to the upper bound
\bea
M_{\chi^0_1} & \le & |\mu| = \left [ \frac{(m_1^2 - m_2^2 \tan^2\beta)}
{(\tan^2\beta - 1)} - \frac{1}{2} M_Z^2 \right]^{\frac{1}{2}}
\label{bound4}
\eea
The importance of this bound stems from the fact that it relates the two 
sectors, namely the supersymmetry breaking gaugino masses and the 
Higgs masses. One can then use the RG 
evolution for the parameters on the RHS of the bounds to evaluate the
bound. Note that the RG equations  for the parameters on the RHS of
the bounds involve the gaugino masses, and will, therefore, 
depend on the boundary conditions for the gaugino masses, and
hence on the different supersymmetry breaking models for the
gaugino mass parameters. 
In Fig. \ref{fig:radcorr} we have plotted the upper limits for
the lightest neutralino mass following from (\ref{bound1})
for the different supersymmetry breaking models as a function of
$m_{\tilde\chi_1^\pm}$.
For all the four  models studied we plot the tree-level masses, 
and the two-loop radiatively corrected masses calculated
using  SOFTSUSY(v.3.0.13)~\cite{Allanach:2001kg}.
In Fig. \ref{fig:radcorr}d, we plot also the upper limit from 
Eq. (\ref{bound4}), which gives the lowest mass upper bound in this
case.  Note that the mass of the lighter of the charginos is close to
the value of the $\mu$-parameter.
The  neutralino masses have been calculated assuming $\tan\beta=10$,
and the other parameters as indicated in the Figure.
We note that all gaugino masses can receive radiative corrections up 
to 20\%, and, thus, difference between tree level and radiatively 
corrected neutralino and chargino masses can be significant in all 
models studied in this paper, although the difference is not explicit
in Fig. \ref{fig:radcorr} due to similar magnitude of correction for both
the neutralino and the chargino.


\section{Sum Rules}
We recall that in the  minimal AMSB model, the mass difference between the
lightest chargino and the lightest neutralino is small.
The close proximity of the lightest neutralino and chargino masses is
a direct consequence of  Eq. (\ref{gmass}), which gives for
the ratios of the gaugino mass parameters
$|M_1|:|M_2|:|M_3|\simeq 2.8:1:7.1$,  after taking into account the next
to leading order radiative corrections and the weak scale threshold
corrections \cite{Gherghetta:1999sw} as in (\ref{anomaly1}).
Thus, the winos are the lightest neutralinos and charginos, and
one would expect that the lightest chargino is only slightly heavier
than the lightest neutralino in all anomaly mediated  models.
It is not feasible to obtain  sum rules for the masses of the 
neutralino states, since the physical neutralino mass matrix is 
a $4\times4$ matrix.
However, from the trace of the squares of the neutralino and chargino 
mass matrices, one obtains a sum rule,  which does not contain the Higgs mixing
parameter $\mu$, but  which is present in the mass matrices. 
The sum rule can be written as 
\bea
2\sum_{i = 1}^2 M_{\tilde \chi^\pm_i}^2 - \sum_{i = 1}^4 M_{\tilde
\chi^0_i}^2
& = & \left[M_2^2 - M_1^2 \right] + 4M_W^2 - 2M_Z^2.
\label{gauginosum}
\eea
\begin{figure}
  \includegraphics[scale=0.5]{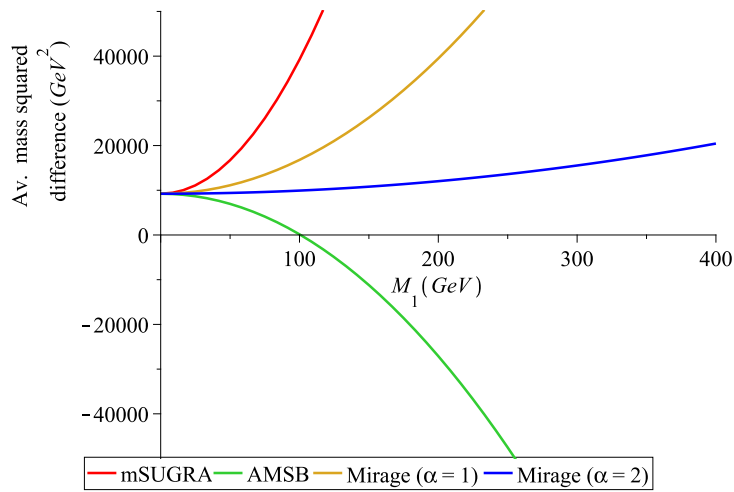}
  \caption{The sum rule (\ref{gauginosum}) plotted for different gaugino
mass patterns.}
  \label{fig:sumrule}
\end{figure}
By using the gaugino mass pattern for a specific model the sum rule
(\ref{gauginosum}) can be expressed as as function of any of the 
gaugino masses.  This is shown in Fig.~\ref{fig:sumrule}.
The average mass difference in the AMSB is first positive, but then
quickly turns negative, while in the minimal SUGRA model it is always
positive. In the mirage mediation model the behavior is determined
by the parameter $\alpha$  with a low value leading to a
mSUGRA-like curve.
Increasing $\alpha$ decreases the gradient of the curve until
$\alpha = 2.17$ (the value for which $M_{1} = M_{2}$) leads to a
constant positive value. We note that this sum rule could be used 
as a signature for different supersymmetry
breaking models, and in the case of mirage mediation it might be
useful for determining the value of $\alpha$,  which can be
calculated from the sum rule for  specific values of  gaugino masses.


\section{Decays of Neutralinos and Charginos}
In this Section we discuss the decays of charginos and
neutralinos in different supersymmetry breaking models that we have 
discussed in this paper. As noted earlier, charginos and neutralinos
are mass eigenstates, which are model-dependent linear combinations
of charged or neutral gauginos and Higgsinos.  Since the mass matrices 
of charginos and neutralinos depend on parameters $M_1$ and 
$M_2$, which are model dependent, the decays will depend on the 
model under consideration. As such the decay
patterns of charginos and neutralinos can serve as a window on the
underlying supersymmetry breaking mechanism in the gaugino sector. 
Here  we shall mostly focus on two-body
tree-level decays of neutralinos and charginos, if they are
kinematically possible.  If the neutralino or 
the chargino is sufficiently heavy, then two-body decays into
a $W, Z^0$, or a Higgs boson and a lighter neutralino or chargino
are the dominant decay modes. Since in supersymmetric models with 
minimal particle content, the lightest Higgs boson is relatively light,
the two-body decay containing the light Higgs boson is expected to be
the dominant decay mode over a large region of parameter space.
However, if some squarks or sleptons are relatively light,
the two body tree-level decays of  a heavy neutralino or chargino
to quark-squark or lepton-slepton  may be important. However, these
decays are phenomenologically less important at a hadron collider 
like LHC, where  neutralinos and charginos
would be produced from the decays of strongly interacting squarks
and gluinos. Neutralinos and charginos, which are heavier than
squarks, would be hard to study at a hadron collider.


We recall that if a heavier $\tilde \chi_i^0~(i = 2, 3, 4)$
or a chargino $\tilde \chi_j^+~(j = 1, 2)$ is produced at  a collider,
it will decay via a cascade until the lightest 
neutralino~( $\tilde \chi_1^0$)  is produced.
Thus, we are here interested in the branching ratios for
the two-body decays
\begin{eqnarray}
&&\tilde \chi_i^0 \rightarrow \tilde \chi_j^0 + Z^0, \quad
\tilde \chi_i^0 \rightarrow \tilde \chi_j^\pm + W^\mp, \quad
\tilde \chi_i^+ \rightarrow \tilde \chi_j^0 + W^+, \quad 
\tilde \chi_i^+ \rightarrow \tilde \chi_j^+ + Z^0, \label{tdecay4}\\
&&\tilde \chi_i^0 \rightarrow \tilde \chi_j^0 + H_k^0, \quad
\tilde \chi_i^0 \rightarrow \tilde \chi_j^\pm + H^\mp, \quad
\tilde \chi_i^+ \rightarrow \tilde \chi_j^0 + H^+,  \quad
\tilde \chi_i^+ \rightarrow \tilde \chi_j^+ + H_k^0. \label{thdecay4}
\end{eqnarray}
These two body decays will dominate any tree-level 
three-body decays mediated by virtual squark or slepton
exchange. These decays will also dominate any two-body decay, which is
forbidden at the tree level, but which can proceed via loops, such as
$\tilde \chi_i^0 \rightarrow \tilde \chi_j^0 + \gamma$.


If some of the neutralinos and charginos are heavier than some of
the squarks and sleptons, then the two-body decays
\bea
\tilde\chi \rightarrow q \, \tilde q, \, \,  l \, \tilde l \label{tsdecay1}
\eea
can compete with the two-body decays into $W, Z^0, H$ discussed above.
The analytical expressions for the 
branching ratios of charginos and neutralinos into 
$W, Z^0$, and Higgs bosons, as well as into squarks/sleptons
for arbitrary neutralino and chargino mixing angles are given 
in~\cite{Gunion:1987yh}.  


For the evaluation of branching ratios of charginos and neutralinos,
we have calculated the spectra of the supersymmetric particles using
SOFTSUSY(v.3.0.13)~\cite{Allanach:2001kg}, and the decays of the
supersymmetric particles using SUSY-HIT(v.1.3 with SDECAY v1.3b/HDECAY
v3.4)~\cite{Djouadi:2006bz}.  In the following analysis we have used
the parameter values $\tan\beta = 10$ and sign$(\mu)=+1$ for all
models. In addition, for the mSUGRA and AMSB models, we have used
$m_0=1$ TeV. For the mirage mediation of supersymmetry breaking, 
we have used $\alpha = 1/2$ and $\alpha = 1$ with $c_{i}=a_{i}=1$.
The mirage mediation scenarios with higher
value for $\alpha$ can lead to squark LSP or tachyons, as will
be discussed in next Section.
While calculating the decay rates of charginos and neutralinos,
we have imposed the experimental constraints following from 
LEP sparticle mass limits and the LEP lower bound on the lightest
Higgs boson mass. Thus, in each of the following figures, the green 
(dark in grayscale) color denotes the area where LEP sparticle mass limits 
are violated,  and light blue color (light) denotes the area where the 
lightest Higgs mass is below 114~GeV. In addition, the vertical 
\textsf{lsp}-denoted line in the mirage
$\alpha=1$ figures indicates the minimum mass for which the neutralino
is the LSP. The \textsf{bsg}-denoted line represents the  $BR(b\to
s\gamma)$ constraint. In the anomaly mediated supersymmetry breaking models,
the \textsf{bsg} line practically coincides with the Higgs mass limit for 
these parameters, and is not
drawn explicitly in order to reduce too many curves in the 
figure. (In the mSUGRA figures
the constraint is obeyed throughout the mass ranges.) The LEP and
$b\to s \gamma$ constraints are calculated with micrOmegas (see Sec.~\ref{sec:RD}).
An asterisk after the decay mode indicates that the charge
conjugated mode is also included in the plotted value. A label of the form
$\Neu 1 \bar x x/\bar y y$ indicates that the plotted value includes
both channels $\Neu 1 \bar x x$ and $\Neu 1 \bar y y$, and that each
channel contributes equally.

We begin by commenting the production of charginos and neutralinos at 
the LHC.
The direct pair production of these particles is rare
\cite{Beenakker:1999xh} compared to the pair production of strongly 
interacting particles \cite{Beenakker:1996ch}.
However, neutralinos and charginos are also produced in the cascade 
decays of squarks and gluinos.
As an example, we study the decays of squarks 
to neutralinos and charginos.

\begin{figure}
  \psfrag{BR}[c][c]{$Br({\tilde u_L}\to \tilde x y)$}
  \psfrag{mD}[c][c]{$m_{\tilde u_L}$ \footnotesize[GeV]}
  \psfrag{neu1u}[r][r]{\scriptsize$\Neu 1 u$}
  \psfrag{neu2u}[r][r]{\scriptsize$\Neu 2 u$}
  \psfrag{neu3u}[r][r]{\scriptsize$\Neu 3 u$}
  \psfrag{neu4u}[r][r]{\scriptsize$\Neu 4 u$}
  \psfrag{cha1d}[r][r]{\scriptsize$\Chap 1 d$}
  \psfrag{cha2d}[r][r]{\scriptsize$\Chap 2 d$}
  \psfrag{gluu}[r][r]{\scriptsize$\tilde g u$}
  \psfrag{neu1t}[r][r]{\scriptsize$\Neu 1 t$}
  \psfrag{aa}{\footnotesize (a)}
  \psfrag{dd}{\footnotesize (d)}
  \psfrag{cc}{\footnotesize (c)}
  \psfrag{bb}{\footnotesize (b)}
  \psfrag{h114}[l][l]{\scriptsize$m_h <114$}
  \includegraphics{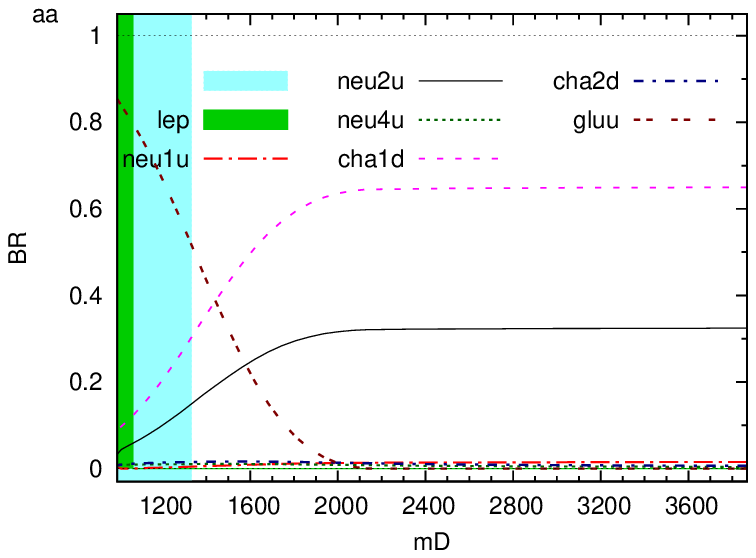}
  \psfrag{h114}[r][r]{\scriptsize$m_h\! <\!114$\! GeV\!}
  \includegraphics{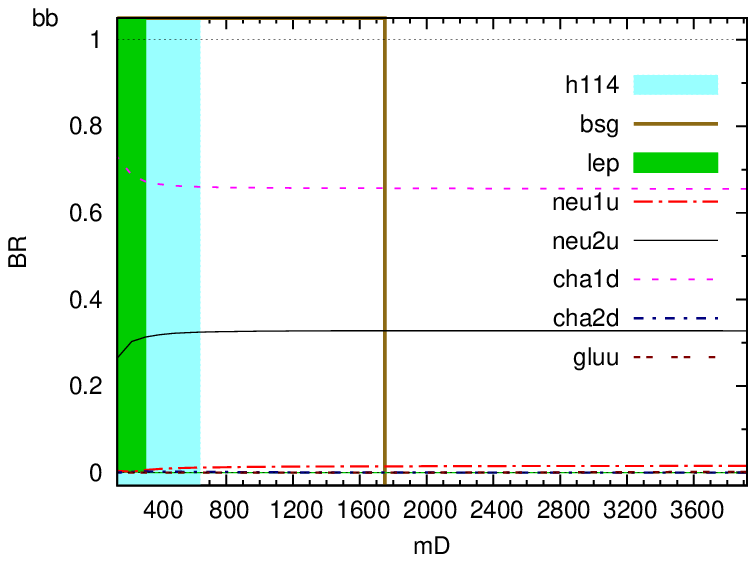}
  \includegraphics{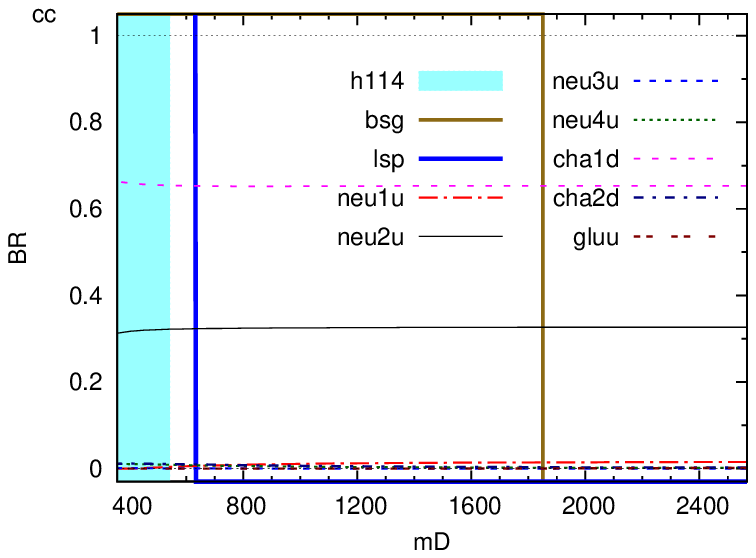}
  \includegraphics{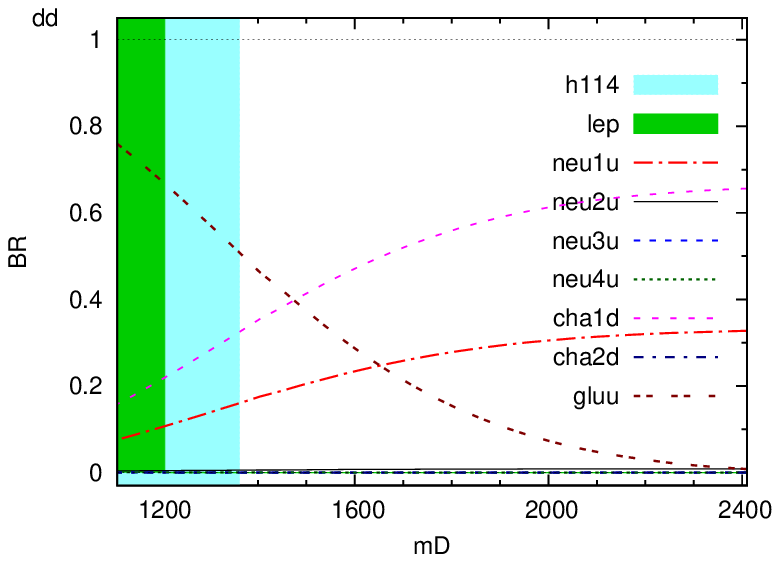}
  \caption{Branching ratios for the two body decays of ${\tilde u_L}$
    to gauginos and Higgsinos in (a) mSUGRA model, (b) mirage
    mediation scenario with $\alpha=0.5$ and (c) $\alpha=1$, and (d)
    the AMSB scenario.  Shadings specify the region where the LEP mass
    limits are not satisfied (dark) or the lightest Higgs mass is
    below 114 GeV (light). See the text for more details.}
  \label{fig:brSUL}
\end{figure}

The decay branching ratio of the left-handed u-squark is depicted 
in Fig.~\ref{fig:brSUL}.
It is seen that if gluino remains lighter than the squark, the
decay is dominantly via the strong coupling to gluino and quark, and then
further gluino would decay to the lightest neutralino with a quark pair 
or a gluon.  
If the two-body decay to quark and gluino is kinematically not
possible, the squark decays mainly to the lighter chargino and 
quark, $\tilde\chi_1^{+} d$, or to the second lightest neutralino
and quark, $\tilde\chi_2^0 u$, in mSUGRA and mirage patterns.
In the AMSB, the branching ratio to $\tilde\chi_1^0 u$ is larger than
to $\tilde\chi_2^0 u$
This is due to the fact that in the mSUGRA and mirage patterns the
neutralino $\chi_2^0$ is mainly wino, while in the AMSB pattern 
$\chi_1^0$ is dominantly wino.
The right-handed squark $\tilde u_R$ decays most of the time to 
$\chi_1^0 u$ in the mSUGRA and mirage patterns, and to $\chi_2^0 u$ in
the AMSB case, since for $\tilde u_R$ the decay to the bino dominated
neutralino is favoured.

\begin{figure}
  \psfrag{BR}[c][c]{$Br({\tilde t_1}\to \tilde x y)$}
  \psfrag{mD}[c][c]{$m_{\tilde t_1}$ \footnotesize[GeV]}
  \psfrag{neu1t}[r][r]{\scriptsize$\Neu 1 t$}
  \psfrag{neu2t}[r][r]{\scriptsize$\Neu 2 t$}
  \psfrag{neu3t}[r][r]{\scriptsize$\Neu 3 t$}
  \psfrag{neu4t}[r][r]{\scriptsize$\Neu 4 t$}
  \psfrag{cha1b}[r][r]{\scriptsize$\Chap 1 b$}
  \psfrag{cha2b}[r][r]{\scriptsize$\Chap 2 b$}
  \psfrag{glut}[r][r]{\scriptsize$\tilde g t$}
  \psfrag{neu1c}[r][r]{\scriptsize$\Neu 1 c$}
  \psfrag{neu1bW}[r][r]{\scriptsize$\Neu 1 b W^{+}$}
  \psfrag{neu1bff}[r][r]{\scriptsize$\Neu 1 b f \bar f'$}
  \psfrag{aa}{\footnotesize (a)}
  \psfrag{dd}{\footnotesize (d)}
  \psfrag{cc}{\footnotesize (c)}
  \psfrag{bb}{\footnotesize (b)}
  \psfrag{h114}[l][l]{\scriptsize$m_h <114$}
  \includegraphics{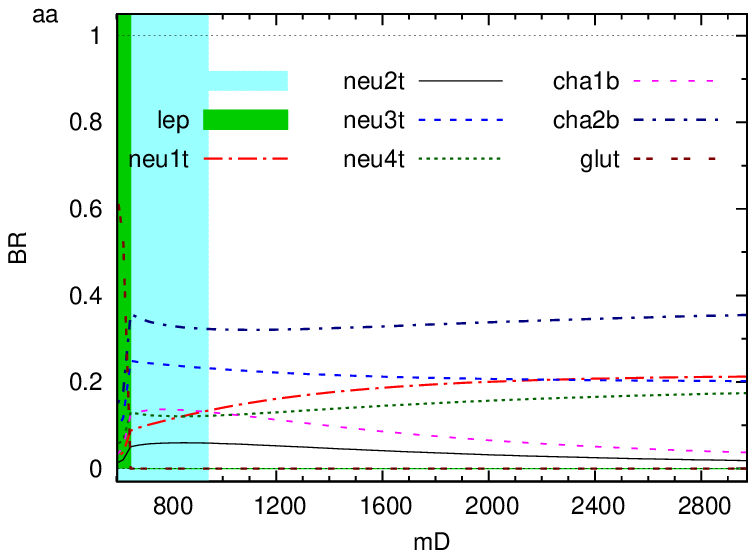}
  \psfrag{h114}[r][r]{\scriptsize$m_h\! <\!114$\! GeV\!}
  \includegraphics{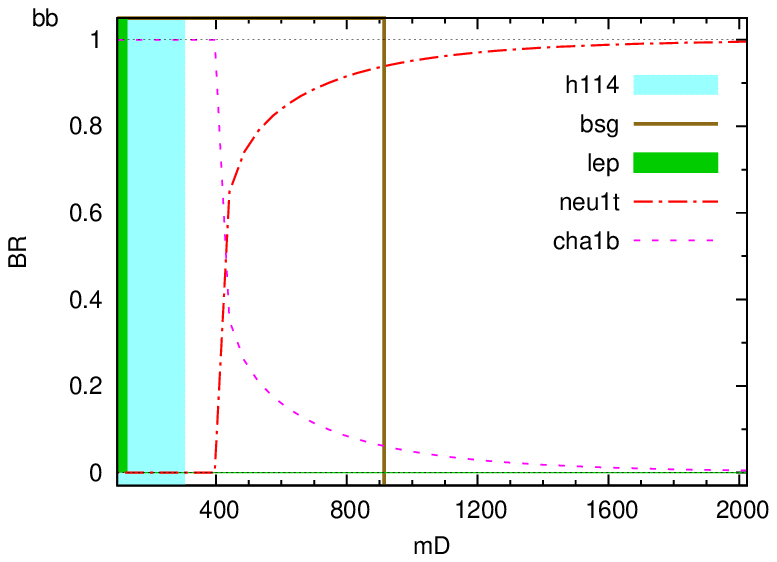}
  \psfrag{h114}[r][r]{\scriptsize$m_h <114$}
  \includegraphics{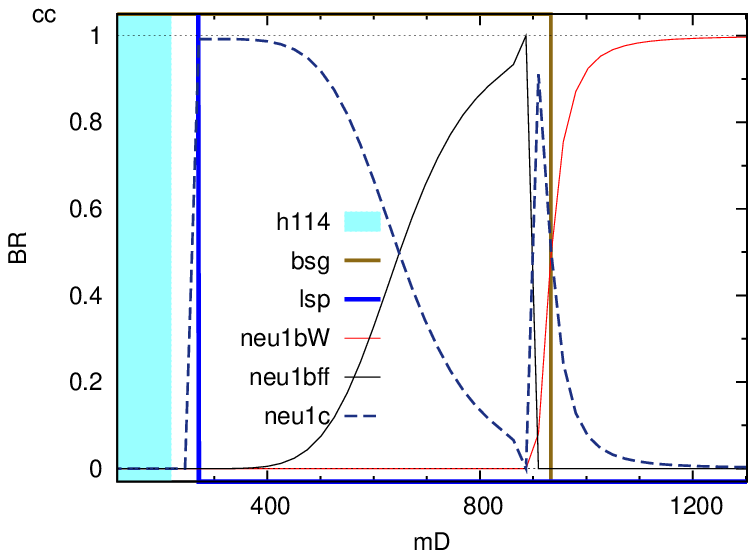}
  \psfrag{h114}[r][r]{\scriptsize$m_h\! <\!114$\! GeV\!}
  \includegraphics{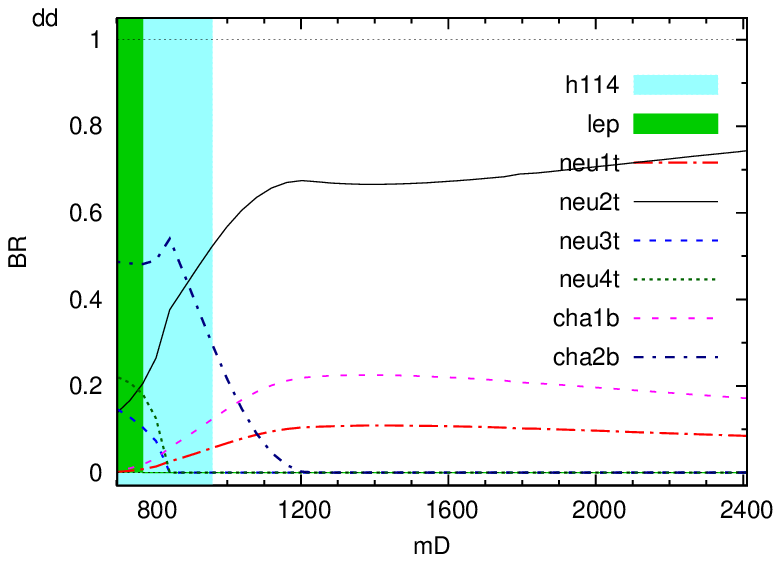}
  \caption{Branching ratios for the largest decay modes of ${\tilde t_1}$
    to gauginos and Higgsinos in (a) mSUGRA model, (b) mirage
    mediation scenario with $\alpha=0.5$ and (c) $\alpha=1$, and (d)
    the AMSB scenario.  Shadings specify the region where the LEP mass
    limits are not satisfied (dark) or the lightest Higgs mass is
    below 114 GeV (light). See the text for more details.}
  \label{fig:brST1}
\end{figure}

For the chosen set of parameters with $m_0=1$ TeV, the squarks are
relatively heavy, especially when the experimentally allowed range 
for $b\rightarrow s\gamma $ is taken into account.
Thus one might want to consider the third generation stops, which are
lighter than the other squarks due to mixing.
In Fig.~\ref{fig:brST1} we have plotted the two-body decay modes of
$\tilde t_1$ to neutralinos, charginos and gluinos as a function of the
stop mass.
One finds that especially in the case of mSUGRA and AMSB,  
the cascades produce heavier neutralinos and charginos,
{\it e.g.} in the AMSB pattern more than 80 \% of the lightest
stop decays to $\tilde\chi_2^{0} t$ or $\tilde\chi_1^{+} b$.
This is different from the two mirage
mediation scenarios, where the lightness of stop causes it to
decay directly to top and the neutralino LSP, or to b-quark, W-boson
and neutralino through a 3-body process.
%
\begin{figure}
 \psfrag{BR}[c][c]{$Br({\Neu 2}\to \tilde x y)$}
 \psfrag{mD}[c][c]{$m_{\Neu 2}$ \footnotesize[GeV]}
  \includegraphics{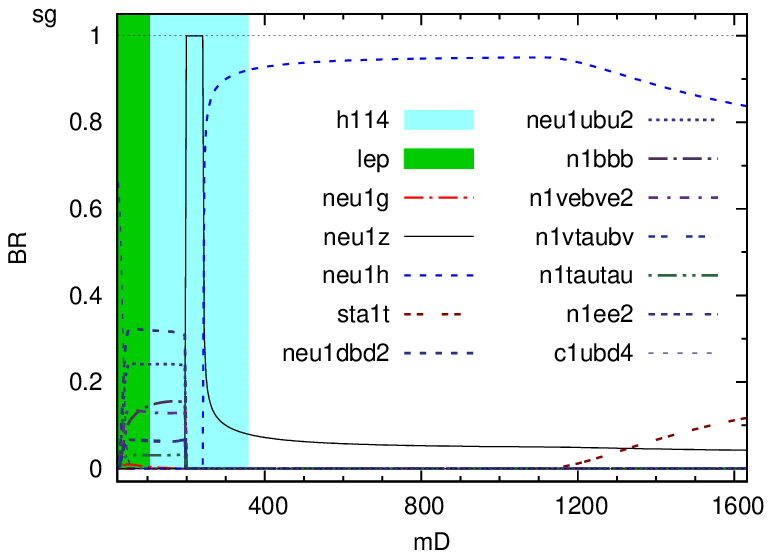}
  \includegraphics{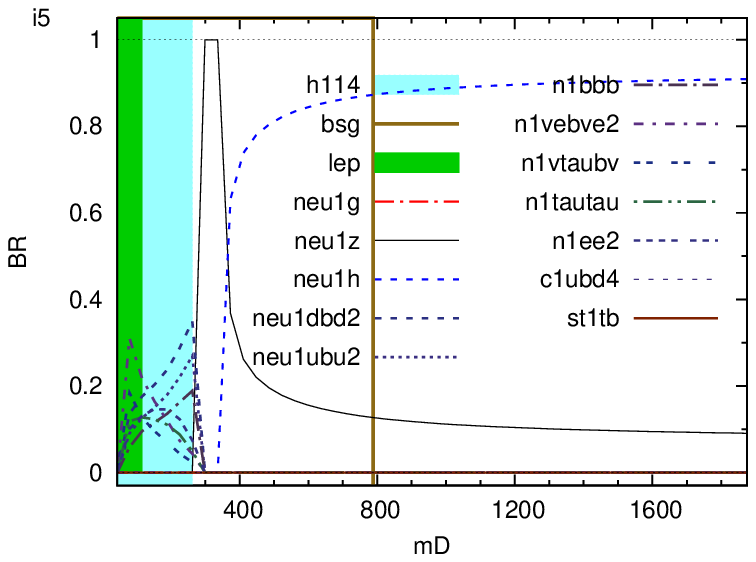}
  \includegraphics{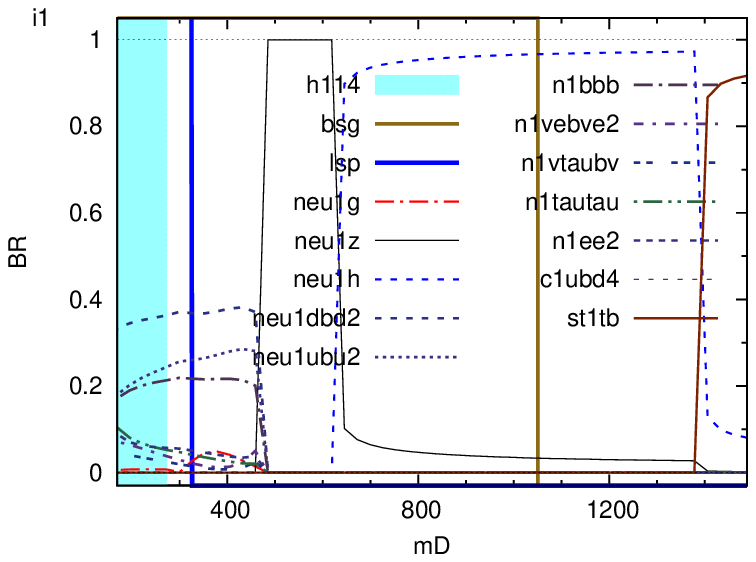}
  \includegraphics{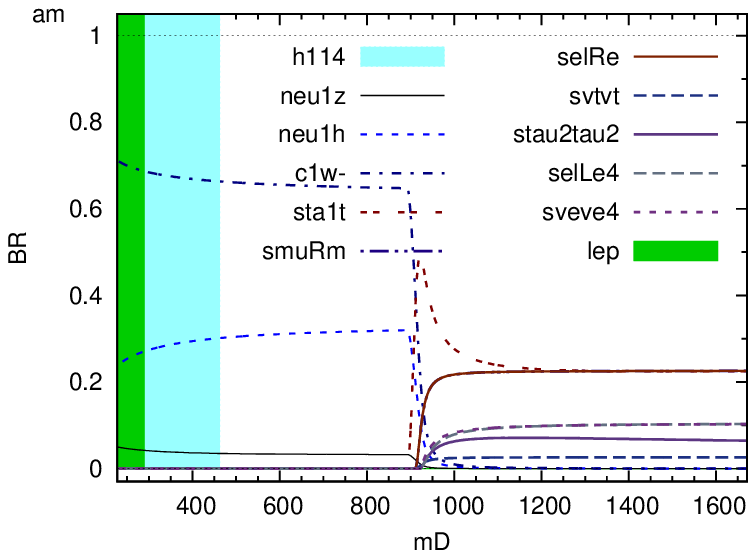}
  \caption{Branching ratios for the two body decays of {\Neu 2} in
    (a) mSUGRA model, (b) mirage mediation scenario with
    $\alpha=0.5$ and (c) $\alpha=1$, and  (d) the AMSB scenario.
    Also the three body channels are shown where no two body
    decays are possible. The $(*)$ in the channel label signifies that
    the channel includes also the charge conjugated final
    state. Shadings specify the region where the LEP mass limits are
    not satisfied (dark) or the lightest Higgs mass is below 114 GeV
    (light).  }
  \label{fig:brN2}
\end{figure}
In Fig.~\ref{fig:brN2} we have plotted the dominant 
decay modes of \Neu 2 as a function of its mass 
for mSUGRA, the mirage mediation models, and the  AMSB, 
respectively. We note from Fig.~\ref{fig:brN2} that 
the dominant mode of \Neu 2 in mSUGRA is into the {\Neu 1} and the
lightest Higgs boson, when kinematically possible, and into {\Neu 1}
and the $Z^0$ boson for a light {\Neu 2}. The same applies for the two
mirage mediation scenarios; one can see the opening of the Higgs mode
to be pushed gradually to higher neutralino masses as the mirage
scenario further deviates from the pure mSUGRA by increasing $\alpha$.
With an increasing $\alpha$, stop becomes lighter and can be the
lightest supersymmetric particle. In the heavier end of the spectrum 
the lightness of stops enables the stop decay mode.
In the AMSB models, where the light chargino and
neutralino are almost degenerate in mass, the decay mode to the
lighter chargino and W-boson dominates the decay to {\Neu 1}
and the lightest Higgs boson, until the sfermion decay modes become
kinematically possible and begin to dominate.
\begin{figure}
 \psfrag{BR}[c][c]{$Br({\Neu 3}\to \tilde x y)$}
 \psfrag{mD}[c][c]{$m_{\Neu 3}$ \footnotesize[GeV]}
  \includegraphics{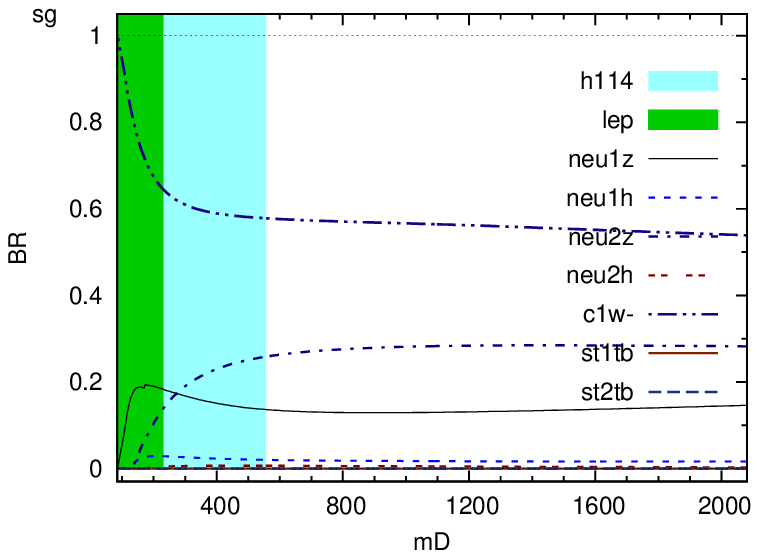}
  \includegraphics{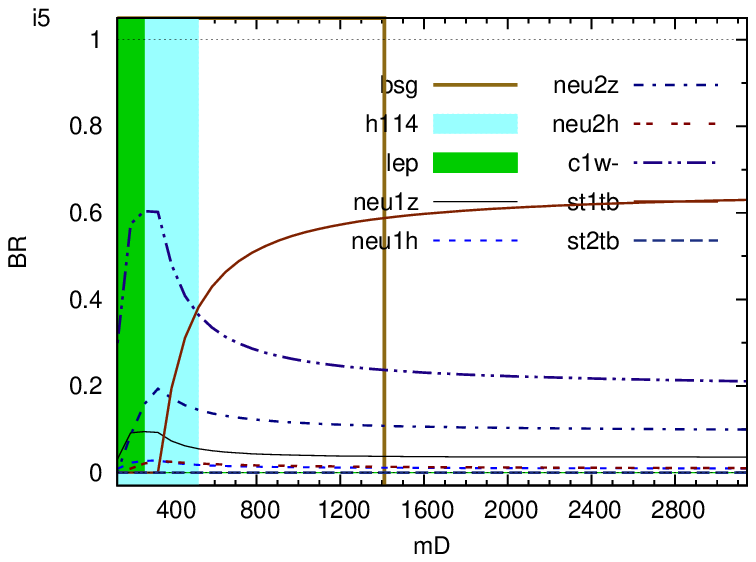}
  \includegraphics{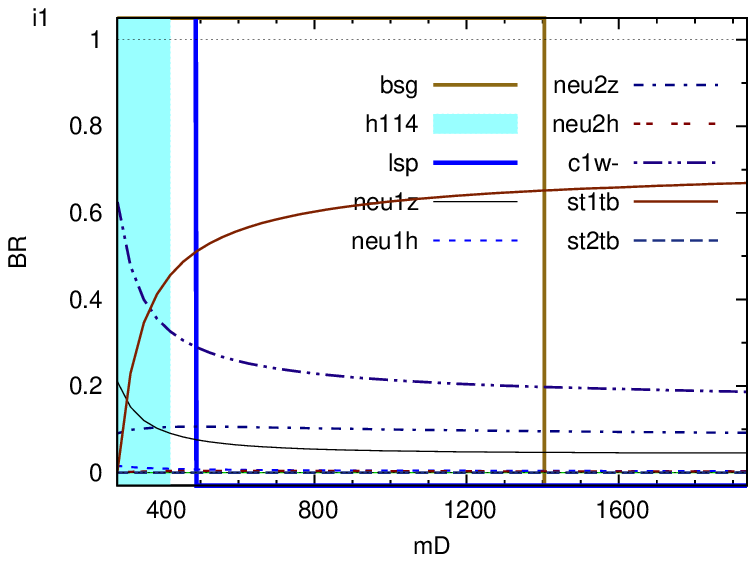}
  \includegraphics{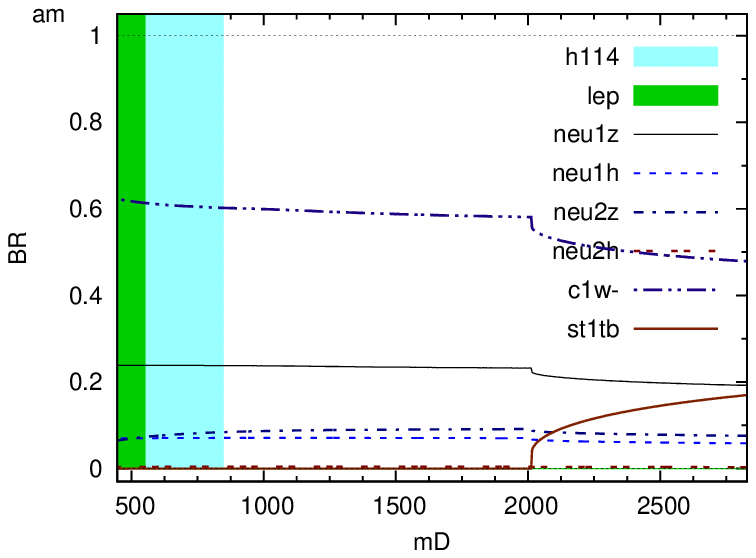}
  \caption{Branching ratios for the decays of {\Neu 3}  
    in  (a) mSUGRA model, (b) mirage mediation scenario with $\alpha=0.5$ 
    (c) in the  mirage mediation scenario with $\alpha=1$,
    and (d) the AMSB scenario.}
  \label{fig:brN3}
\end{figure}

In Fig.~\ref{fig:brN3} we have plotted the dominant two body decay
modes of the \Neu 3 as a function of its mass for mSUGRA, the mirage
mediation models, and the AMSB, respectively. For \Neu 3 the chargino-$W$
mode is available in mSUGRA, and it dominates.  However, the 
$\Neu 2 Z^0$ mode starts competing with increasing mass of \Neu 3.
In the mirage mediation models the stop-top channel is available 
in addition to the $\Cha 1 W$ and the $Z^0$ decay  modes.
Furthermore, in the mirage mediation scenario, the light stop-top 
channel takes the role of the dominant decay mode for a large mass range, 
where the two mirage scenarios give quite similar branching ratios for 
the \Neu 3 decay.
This is in contrast to the AMSB scenario, where the $\Cha 1 W$ and
$\Neu 1 Z^0$ modes dominate.
\begin{figure}
 \psfrag{BR}[c][c]{$Br({\Neu 4}\to \tilde x y)$}
 \psfrag{mD}[c][c]{$m_{\Neu 4}$ \footnotesize[GeV]}
  \includegraphics{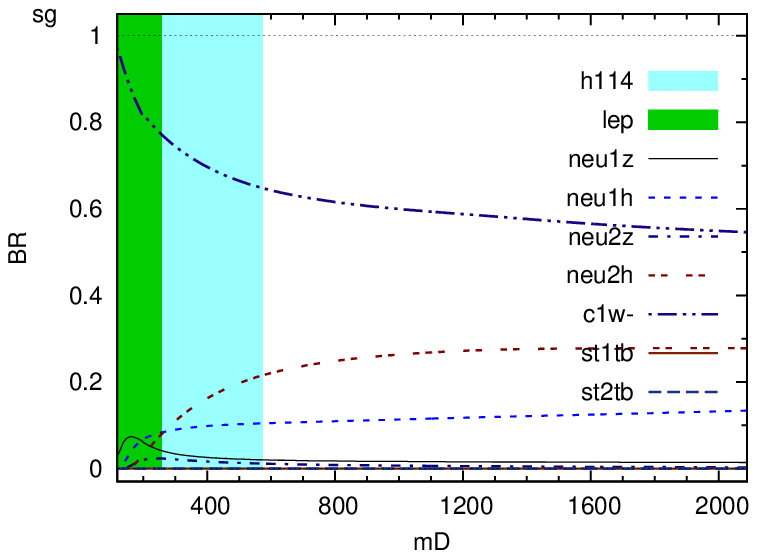}
  \includegraphics{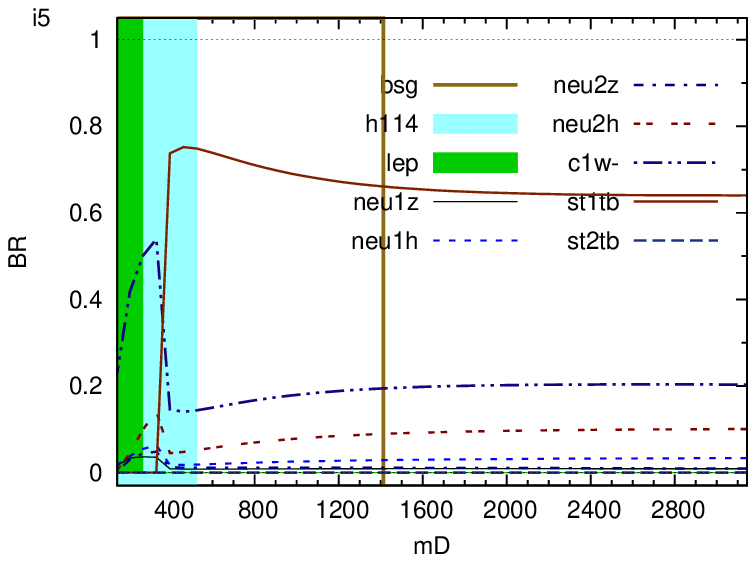}
  \includegraphics{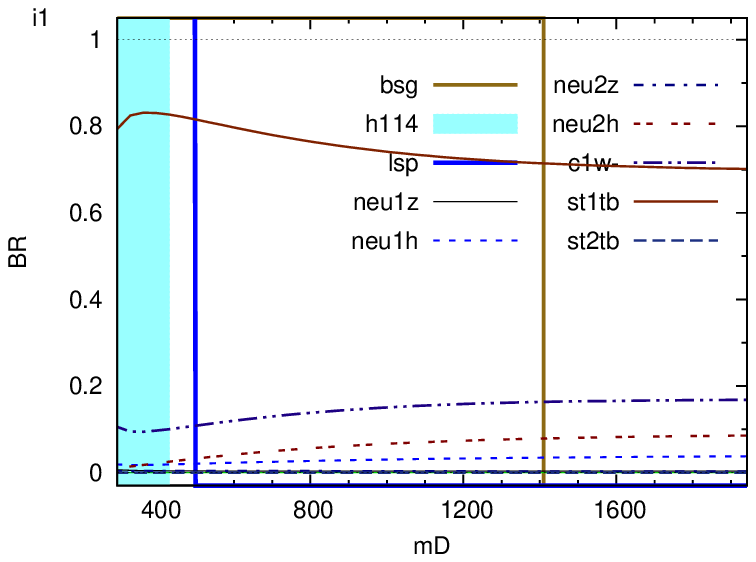}
  \includegraphics{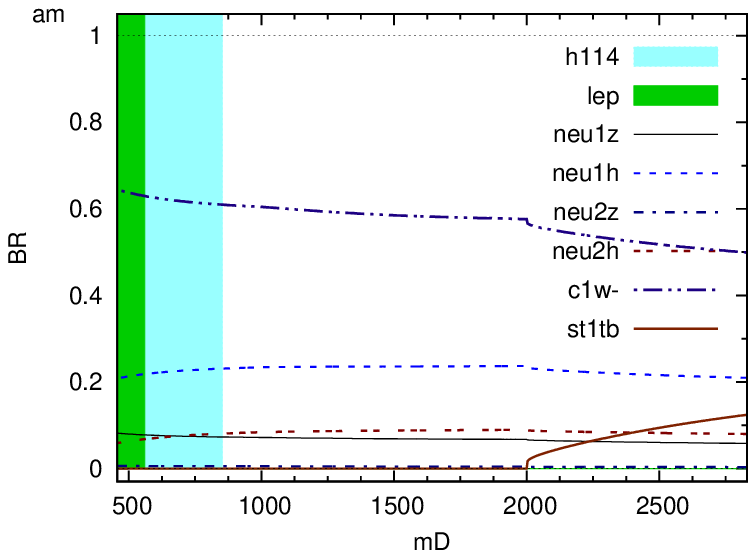}
  \caption{Branching ratios for the decays of {\Neu 4}  
    in  (a) mSUGRA model, 
    (b) mirage mediation scenario with $\alpha=0.5$ 
    and (c) $\alpha=1$,
    and (d) the AMSB scenario.}
  \label{fig:brN4}
\end{figure}

In Fig.~\ref{fig:brN4} we have plotted the dominant two body decay
modes of the \Neu 4 as a function of mass of \Neu 4.  In mSUGRA and
AMSB scenarios the $\Cha 1 W$ mode dominates, whereas in the mirage
mediation the stop-top channel has the largest branching fraction. For
the relevant mass range, in the two mirage scenarios, the \Neu 4
decays look similar. Furthermore, the light Higgs decay modes are
present in each breaking scenario.
\begin{figure}
 \psfrag{BR}[c][c]{$Br({\Chap 1}\to \tilde x y)$}
 \psfrag{mD}[c][c]{$m_{\Chap 1}$ \footnotesize[GeV]}
  \includegraphics{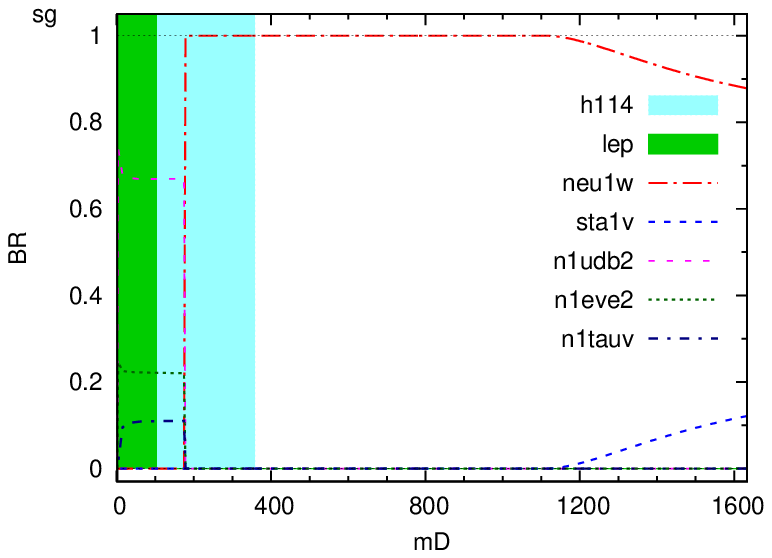}
  \includegraphics{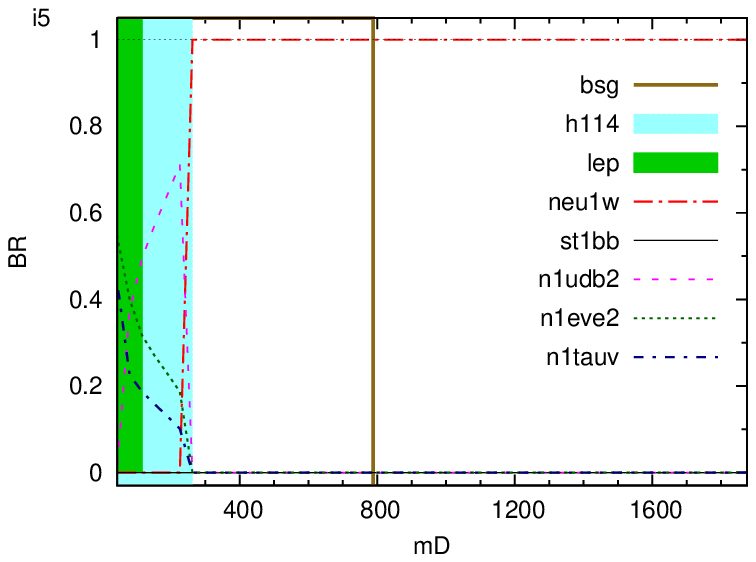}
  \includegraphics{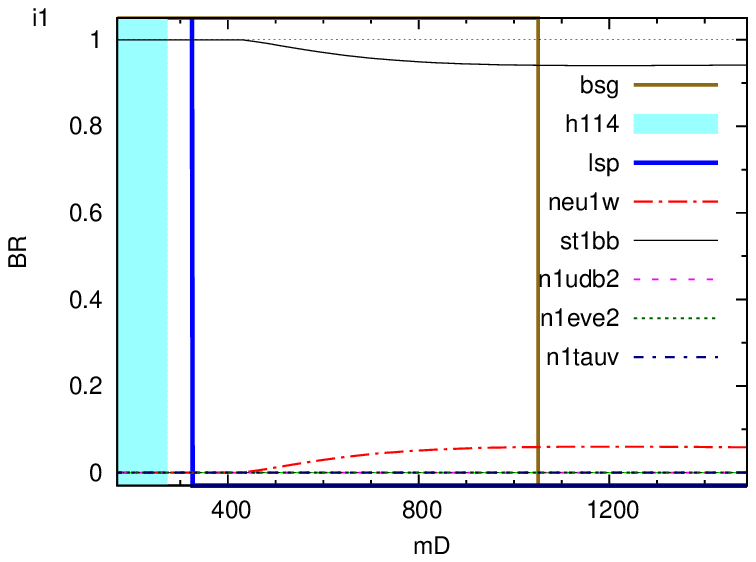}
  \includegraphics{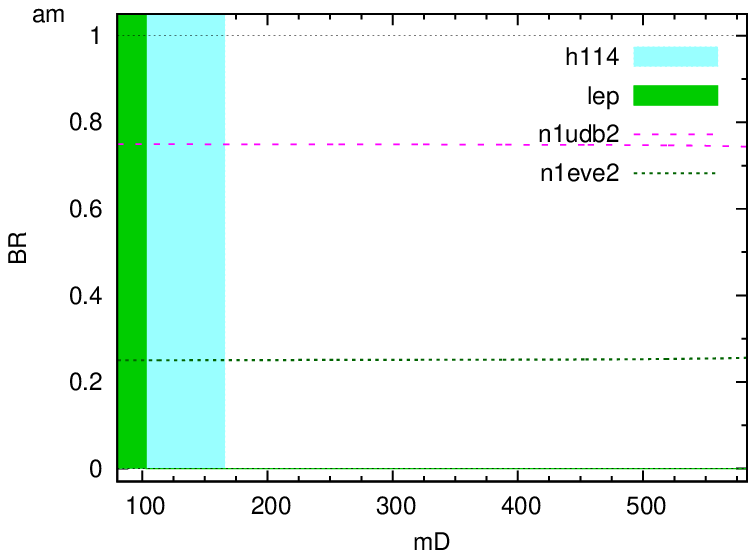}
  \caption{Branching ratios for the decays of {\Chap 1}  
    in  (a) mSUGRA model, (b) mirage mediation scenario with $\alpha=0.5$ 
    and (c) $\alpha=1$, and (d) the AMSB scenario.}
  \label{fig:brC1}
\end{figure}

In Fig.~\ref{fig:brC1} the dominant decay modes of \Cha 1
are plotted as a function of its mass for the
different supersymmetry breaking models. 
In mSUGRA the only possible two-body decay mode,  until the slepton
(stau) channel becomes available, is the $\Neu 1 W$ mode.
In the mirage mediation scenario the lightness of the stop enables the
$\tilde t_1 \bar b$ mode, which becomes the dominant one for
$\alpha=1$. The other available channel is the $\Neu 1 W$ mode, which
in fact is the only two-body mode in the $\alpha=0.5$ case, thus
making the two mirage scenarios distinguishable with respect to the
lighter chargino decay. Also, the neutralino LSP and the $b\to
s\gamma$ requirements push
the chargino mass to be reasonably heavy in the $\alpha=1$ mirage scenario.
In the AMSB models the lighter chargino mass is so close to the
lightest neutralino mass that only three body decays are available.
\begin{figure}
 \psfrag{BR}[c][c]{$Br({\Chap 2}\to \tilde x y)$}
 \psfrag{mD}[c][c]{$m_{\Chap 2}$ \footnotesize[GeV]}
  \includegraphics{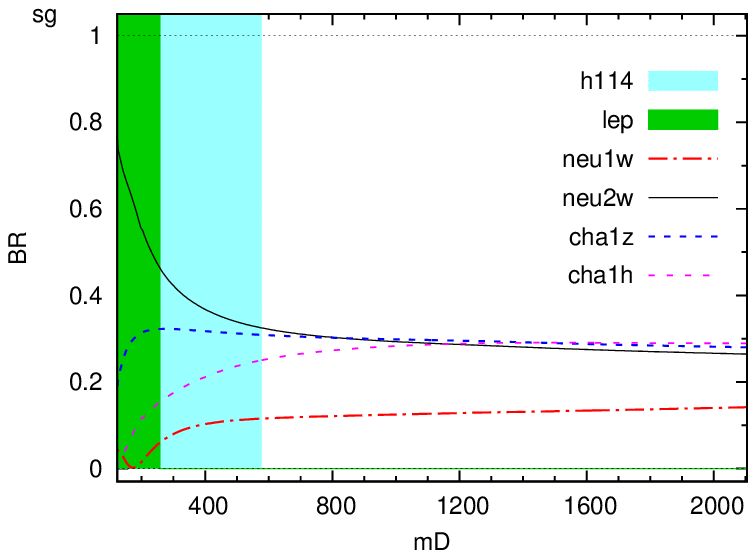}
  \includegraphics{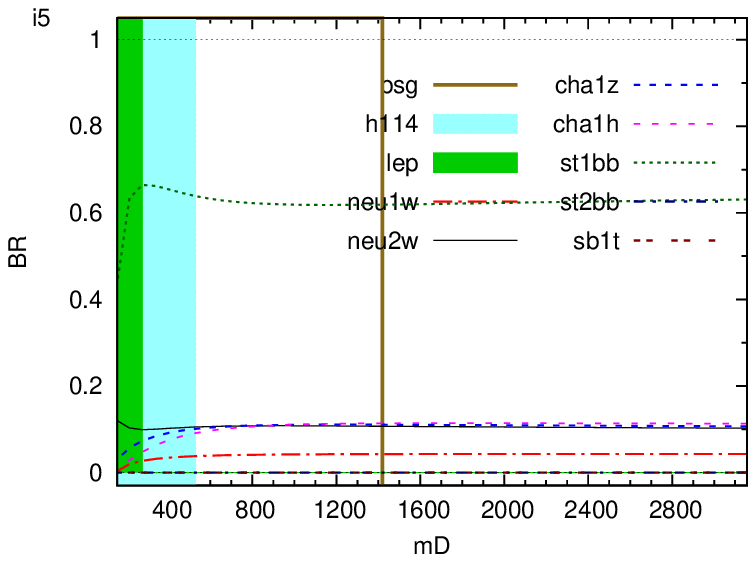}
  \includegraphics{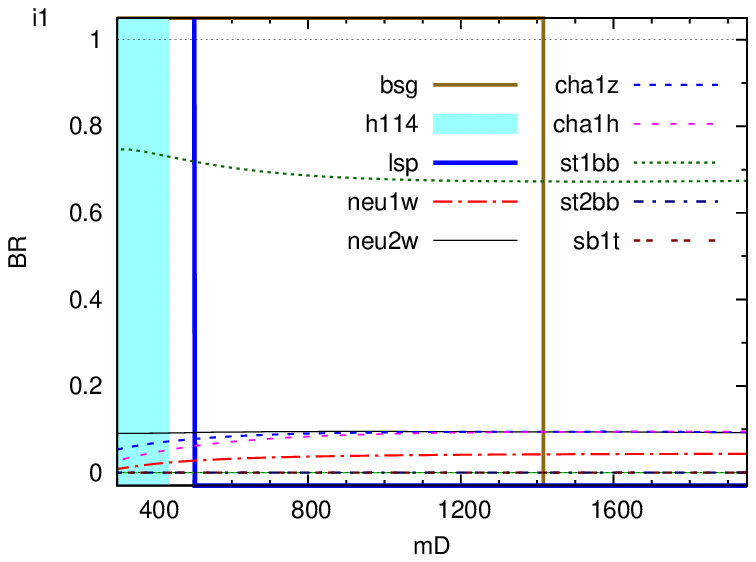}
  \includegraphics{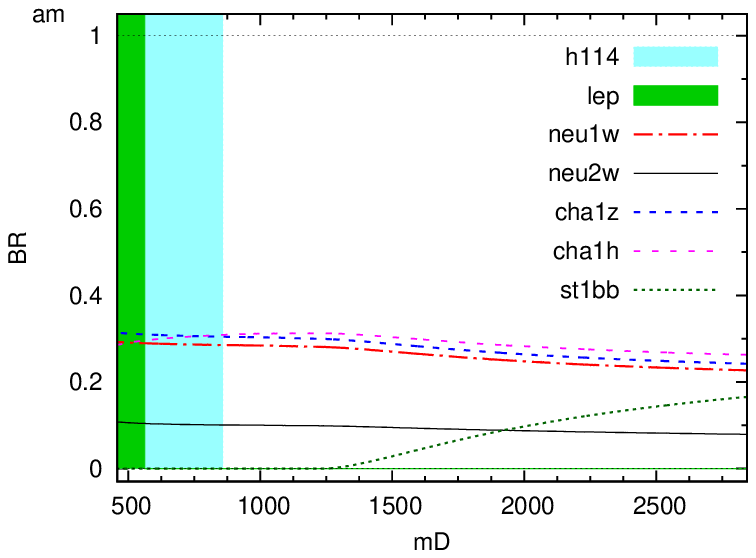}
  \caption{Branching ratios for the decays of {\Chap 2}  
    in  (a) mSUGRA model, 
    (b) mirage mediation scenario with $\alpha=0.5$ 
    and (c) $\alpha=1$, and (d) the AMSB scenario.}
  \label{fig:brC2}
\end{figure}

Finally, in Fig.~\ref{fig:brC2} we plot the dominant two body decay
modes of the \Cha 2 as a function of its mass for the models
discussed in this paper.  In mSUGRA the $W$ and $Z^0$ modes dominate 
for the low mass range, but eventually the light Higgs channel becomes 
a competing mode.
In the AMSB scenario the light Higgs, $Z^0$ and $W$ modes have approximately
the same branching ratio.  In contrast to the other models, in the mirage 
mediation the $\tilde t_1 \bar b$ mode dominates by a large margin.


In general, it is obvious that in the AMSB models the close proximity
of the lightest neutralino and the lighter chargino is important in
the decays of heavier neutralinos and chargino as well.  In addition
to forcing the lighter chargino to decay via a three body mode, this
feature makes the second lightest neutralino to decay dominantly into
the lighter chargino and $W$ boson or other charged sparticles. The
predictions of the anomaly mediated supersymmetry breaking and the
mSUGRA are very different here.


In mirage mediation the mass spectrum for neutralinos and charginos is
more tightly packed, and thus for $\chi_2^0$ and $\chi_1^\pm$ it is
expected that they decay via a three body mode, if the mass scale is
light enough, which again is a distinguishing feature as compared to
the mSUGRA and AMSB models. The lightness of stop  allows in many
cases the stop decay channel in the mirage mediation models, which
then becomes the dominant decay mode.


\section{Relic Density\label{sec:RD}}

In this Section, we study the implications of different patterns of
supersymmetry breaking gaugino masses for the relic density of
lightest neutralino, and the constraints imposed on the parameter space by
the precise limits on the relic density obtained by the Wilkinson
Microwave Anisotropy Probe (WMAP) satellite. Requiring the lightest
neutralino to form all of the dark matter as a thermal relic is very
limiting constraint on the parameter space, and it should be kept in
mind that the possible dark matter might also be created non-thermally
or the excess thermal production diluted for example by an entropy
increase after the freeze-out
\cite{Kamionkowski:1990ni,Kaplinghat:2000jj,Baltz:2001rq}.  Therefore
we refer to the WMAP constrained parameter space as a WMAP-preferred
relic density area.

\begin{figure}
  \psfrag{tanb}[c][c]{$\tan\beta$}
  \psfrag{mch}[c][c]{$m_{\tilde\chi_1^\pm}$ \scriptsize [GeV]}
  \psfrag{3-body}[l][l]{\scriptsize 3-body}
\psfrag{sta1t}[r][r]{\scriptsize$\tilde \tau_1 \bar\tau (*)$}
\psfrag{selRe}[r][r]{\scriptsize$\tilde e_R \bar e /\tilde \mu_R \bar \mu
  (*)$}
  \centering
  \includegraphics[width=0.45\textwidth]{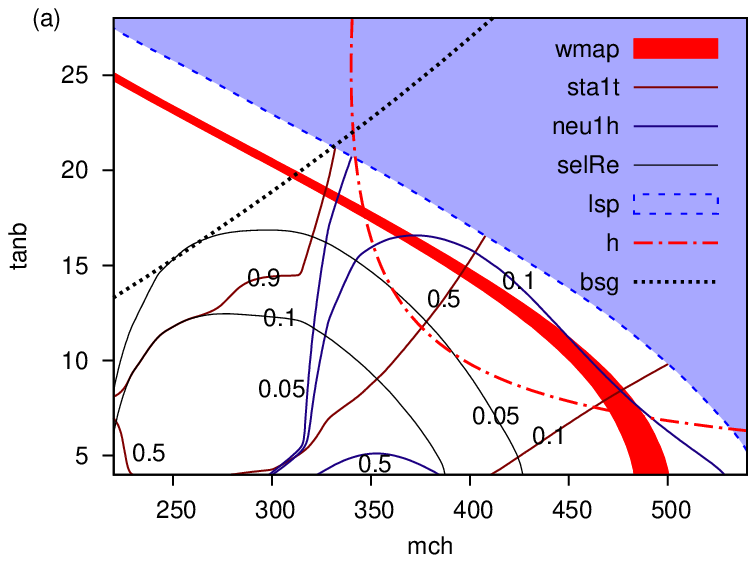}
  \includegraphics[width=0.45\textwidth]{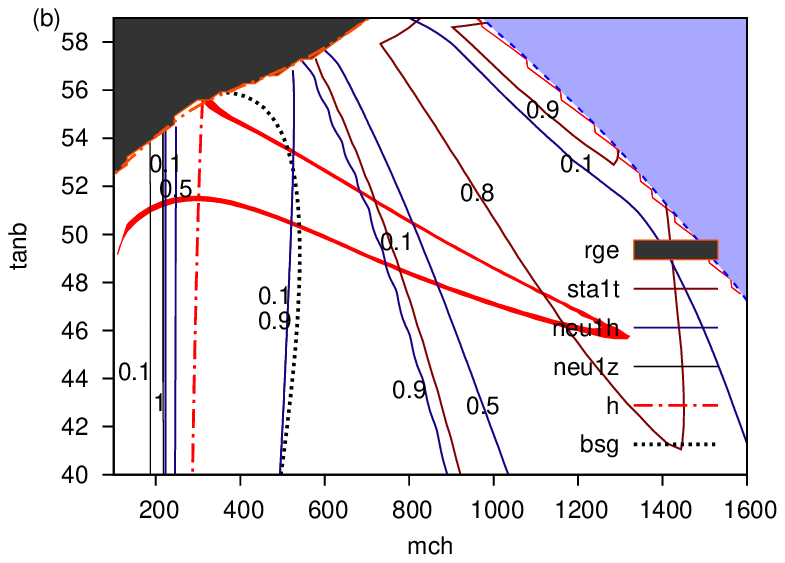}\\
  \includegraphics[width=0.45\textwidth]{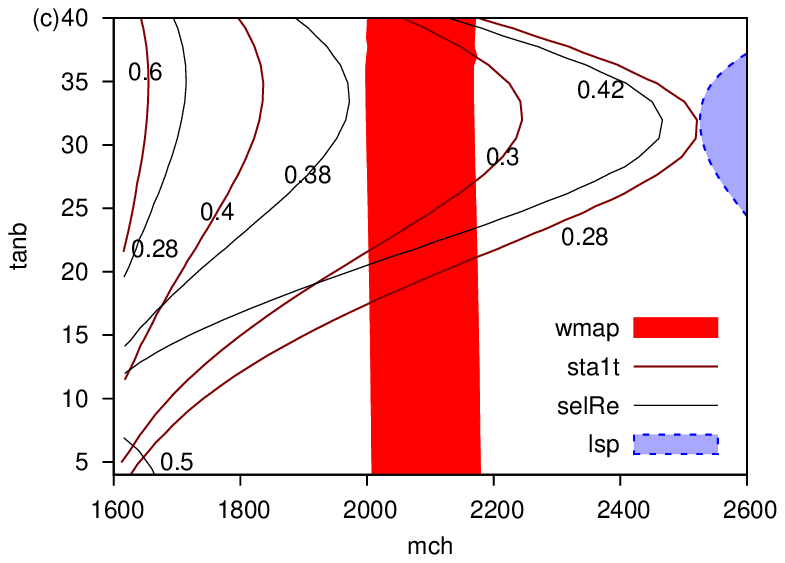}
  \includegraphics[width=0.45\textwidth]{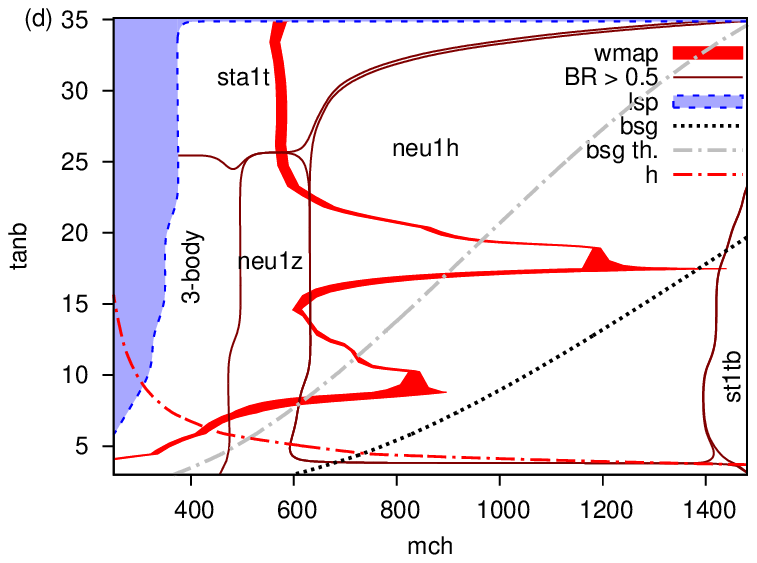}
  \caption{Contours of constant branching ratio for the leading
    two-body decay modes of \Neu 2 superposed on the same plot with
    several constraints for (a) mSUGRA scenario for $m_0=120$ GeV, (b)
    $m_0=1$ TeV, (c) AMSB for $m_0=5$ TeV and (d) the mirage
    mediation for $\alpha=1$. The $b\to s \gamma$ constraint is obeyed
    right of the dotted \textsf{bsg}-denoted line and the lightest Higgs mass is
    more than 114 GeV on the right of the h denoted dash-dotted line.  }
  \label{fig:brAllTanb}
\end{figure}

The relic density in the
mSUGRA \cite{Drees:1992am,Baer:1995nc,Edsjo:1997bg,Ellis:1998kh,
Arnowitt:2001ca,Ellis:2003cw,Baer:2003wx} and AMSB
\cite{Gherghetta:1999sw,Moroi:1999zb,Barr:2002ex,Profumo:2004ex,
Hooper:2003ka}
scenarios has been studied extensively.  Neutralino dark matter in the
mirage mediation SUSY breaking model has been considered in
\cite{Choi:2006im,Nagai:2007ud,Abe:2007je,Baer:2007eh}.
Here we consider the combined information from the relic density and
decay modes of the second lightest neutralino.
In Fig.~\ref{fig:brAllTanb} the main decay modes of \Neu 2
and the WMAP-preferred relic density stripe in the ($M_{\Cha 1},
\tan\beta$) plane are plotted for mSUGRA ($A_0=0, m_0=120$ GeV and
$m_0=1$ TeV), AMSB ($m_0=5$ TeV), and mirage mediation ($\alpha=1$
with $a_i=c_i=1$) for sign$(\mu)=+1$.
The spectrum is calculated using SOFTSUSY
(v.3.0.13)~\cite{Allanach:2001kg}, and the relic density and
constraints using micrOmegas (v.2.2.CPC.i)~\cite{Belanger:2006is}.
For the relic density, we use the WMAP combined three year limits
\cite{Spergel:2006hy}
\begin{eqnarray}
  \Omega_{CDM} h^2 = 0.11054^{+0.00976}_{-0.00956} \quad (2\sigma).
\end{eqnarray}
In the figures below, the filling denoted by \textsf{wmap} is the WMAP
preferred region.  For the $b\to s\gamma$ experimental branching
fraction, the two sigma world average has been used
\cite{Barberio:2007cr},
 $ BR(b\to s \gamma) = (355 \pm 24^{+9}_{-10} \pm 3) \times 10^{-6}$,
and for the particle masses the limits of \cite{Belanger:2006is} are
applied.  In the figures, \textsf{lep} shows an area where the
experimental sparticle mass limits are not met, \textsf{rge} shows an area 
where there are tachyons or no radiative EWSB, and \textsf{lsp} the area
where neutralino is not the LSP.  The curve $m_h=114$ GeV is depicted
in the figures for the reference (dash-dotted line denoted by
\textsf{h}).


In the mSUGRA scenario the WMAP preferred relic density area can be
obtained either for a small $m_0$ with moderate $\tan\beta$, where the
scalars are light and stau coannihilates with light neutralino, or for
a larger $m_0$ with quite large $\tan\beta$, where \Neu 1 annihilates
through Higgs channel. The lightest neutralino is very bino-like, which in
general results in a large relic density.
In Fig.~\ref{fig:brAllTanb}a the WMAP preferred region is shown for
$m_0=120$ GeV.  This is achieved via coannihilation with stau for
$m_{\Cha 1}<500$ GeV and $\tan\beta < 20$.  The dominant decay mode
along the WMAP stripe is to $\tilde{\tau}_{1} \tau$ -pair until
various other leptonic modes and the light Higgs mode take over with
increasing \Cha 1 mass.
While the stau coannihilation region exists also for the heavier bino
masses, the WMAP preferred region becomes very narrow due to the relic
density mass dependence on the LSP mass.
Increasing $m_0$ to 200 GeV would raise the WMAP preferred region to 
around $m_{\Cha 1}\sim 500-700$ GeV and $12<\tan\beta <25$, but the
stripe would be clearly narrower due to stau coannihilation with 
heavier bino.  For $m_0=1$ TeV, $A_0=0$, and $\mu>0$, the dark matter
area can be found around $\tan\beta \simeq 50$ \cite{Baer:2008ih}, as
shown in Fig.~\ref{fig:brAllTanb}b.  Here both neutralino annihilation
through a heavy Higgs resonance and stau coannihilation are effective.
Along the WMAP stripe $\Neu 1 h$ is the dominant decay mode until
$\tilde{\tau}_1 \tau$ becomes kinematically accessible.


We note that since in the AMSB scenario the lightest neutralino is
almost a pure wino for a large region of the parameter space, the
relic density for moderate $m_0$ values tends to be smaller than
the WMAP observation.
Therefore a spectrum with rather large LSP mass is required for
the AMSB to provide the observed relic density.
In Fig.~\ref{fig:brAllTanb}c the value of  $m_0=5$ TeV has been used.  The
WMAP region occurs for $m_{\chi^\pm_1}\simeq 2$ TeV, where
the decay modes to slepton-lepton pair dominate the {\Neu 2}-decays.


In mirage mediation scenario (Fig.~\ref{fig:brAllTanb}d) with
$\alpha=1$ with the mentioned choice of parameters the lightest
neutralino is mostly bino-like. Therefore the existence of the
WMAP-preferred relic density region in general relies on the
coannihilation with the NSLP, which for the large part of the
parameter space is stop.  (For high $\tan\beta$ the NLSP near the
\textsf{lsp}-denoted region is stau.)  Taking into account the $b\to s \gamma$
constraint and requiring that the relic density agrees with the WMAP
observations practically removes the WMAP-preferred region. If a ten
percent error in the theoretical calculation of the $b\to s \gamma$
constraint is assumed, the WMAP-stripe on the right of the \textsf{bsg
th.}-denoted line becomes allowed. The peaking behavior in the relic
density is due to the efficient s-channel annihilation through heavy
Higgs bosons. Around $\tan\beta=10$ the annihilation rate is enhanced
via stops decaying through the Higgs resonance, while around
$\tan\beta=18$ it is the lightest neutralino which annihilates
directly though the resonance.
For a large region of the parameter space the NLSP is the lighter stop, 
so the WMAP-preferred relic density parameter space
follows the neutralino co-annihilation region with stop.
The figure is divided into areas where the dominant \Neu 2 decay mode
exceeds 50 \%. If the constraints are obeyed, the
dominant decay mode for \Neu 2 is to $\chi_1^0 h$ or possibly to
$\tilde t_1 \bar t (*)$ for the WMAP-preferred relic density area.
%
%
\begin{figure}
\psfrag{m2}[c][c]{$M_{3/2}$ \footnotesize[GeV]}
\psfrag{m32}[c][c]{$M_{3/2}$ \footnotesize[GeV]}
  \psfrag{3-body}[l][l]{\scriptsize 3-body}
\psfrag{neu1zb}[r][r]{\scriptsize$\Neu 1 Z$}
\psfrag{neu1hb}[r][r]{\scriptsize$\Neu 1 h$}
\psfrag{sta1tb}[r][r]{\scriptsize$\tilde \tau_1 \bar\tau (*)$}
\psfrag{a6}[r][r]{(a)}
\psfrag{b6}[r][r]{(b)}
\psfrag{alpha}{$\alpha$}
  \centering
  \includegraphics[height=70mm,width=0.47\textwidth]{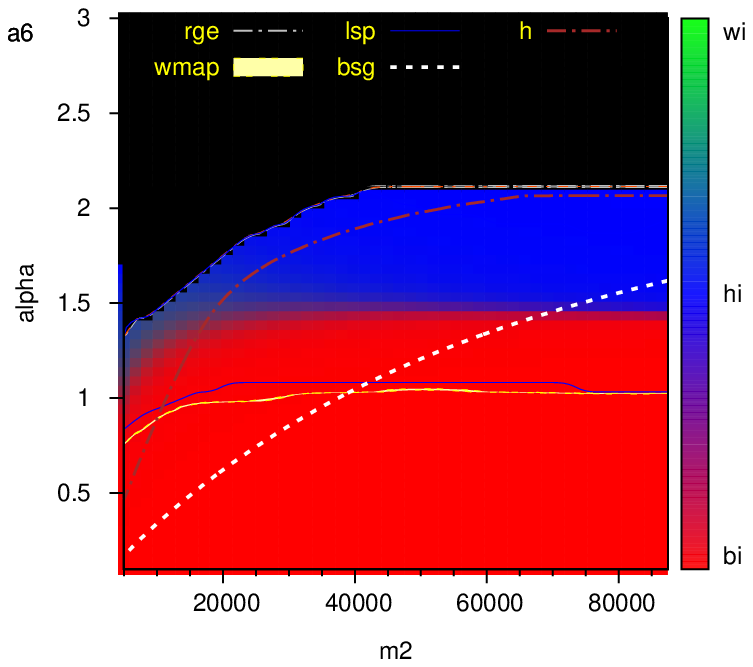}
  \includegraphics[height=70mm,width=0.43\textwidth]{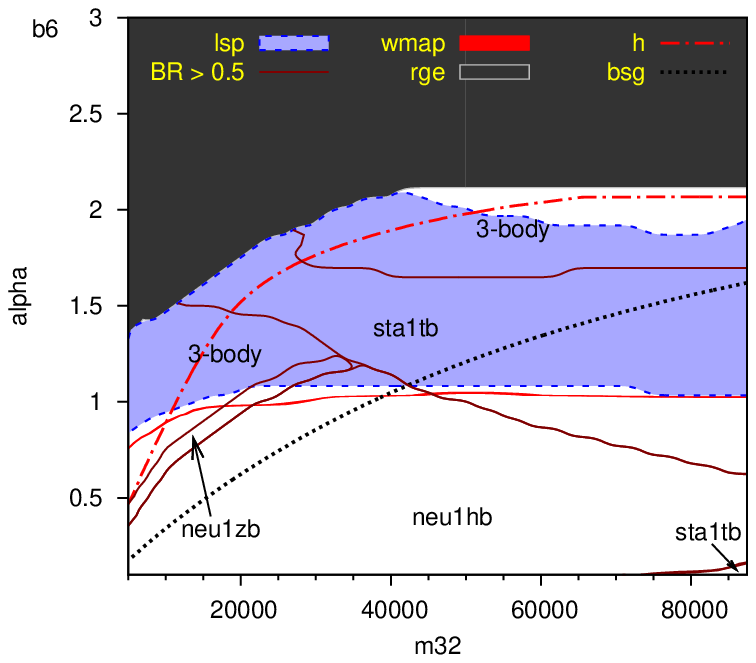}
  \caption{Lightest neutralino composition (a) and the leading \Neu 2
    decay modes (b) in the mirage mediation scenario in ($M_{3/2},
    \alpha$) plane for $\mathrm{sgn} (\mu) = +1,\ \tan\beta=10$ and
    $a_i=c_i=1$. The narrow light yellow band in (a) (red in (b)) indicates the
    WMAP preferred relic density area. The $b\to s \gamma$ constraint
    is obeyed below the dotted \textsf{bsg}-denoted line and the lightest Higgs
    mass is more than 114 GeV below the h denoted dash-dotted
    line. The \textsf{lsp} denoted (light blue) line near the WMAP filling
    limits the area, above which the lightest neutralino is not the
    LSP except for the area near $\alpha=2$, which can better be seen
    in (b). The black areas limited by the \textsf{rge}-denoted line depicts
    the area where there are either tachyons or no REWSB.  In (b) the
    domains of branching ratio exceeding 50 \% for the leading decay modes of
    \Neu 2 are drawn for the same parameters, including the
    constraints.}
  \label{fig:ncMirage}
\end{figure}

In Fig.~\ref{fig:ncMirage}a the composition of the lightest neutralino
and the WMAP-preferred relic density stripe in the mirage mediation
scenario in the ($M_{3/2}, \alpha$) plane are plotted (for
$\tan\beta=10$ and sign$(\mu)=+1$) with $a_i=c_i=1$.  
The lightest neutralino composition is painted in RGB-color encoding
({\it i.e.}, colors (or hues of gray) indicate the particle
as shown in the figure color bar; therefore the mixture of the colors
(or the shading in black and white) describes the nature of the $\Neu
1$-composition).
As the $\alpha$-parameter
increases, the lightest neutralino composition makes a transition from
the bino-like to very Higgsino-like. Above the
\textsf{lsp}-denoted line the stop is the LSP, and therefore that area is not
suitable for the dark matter considerations. However, at sufficiently
high alpha value the REWSB forces the $\mu$-parameter to very light
value, hence reducing the Higgsino like neutralino mass below the
sparticle masses and the lightest neutralino becomes the LSP
again. The areas can  be seen in Fig.~\ref{fig:ncMirage}b, where also
the domains of \Neu 2 decay modes exceeding 50 \% are marked.
The thin WMAP-preferred relic density band lies along the line where
the lightest neutralino can coannihilate with the lighter stop, hence
reducing the relic density to an acceptable level.  Here the \Neu 2
decays dominantly to $\chi_1^0 h$ or to $\tilde{\tau}_1  \tau$ pair,
as indicated in the figure.


\section{Summary and discussion}

The lightest neutralino is expected to be the lightest supersymmetric particle
in supersymmetric models with $R$-parity conservation. It is expected to be
the end product of decays of supersymmetric partners of the Standard Model
particles that are likely to be produced at the Large Hadron Collider.
Thus, its mass, and its properties  are of considerable importance
for the supersymmetric phenomenology. The mass of the lightest neutralino,
as well as those of its  heavier partners, depend on the mechanism of  
supersymmetry breaking in the gaugino sector.


We have carried out a detailed study of the
spectrum of neutralinos and charginos in different models of 
supersymmetry breaking gaugino masses. Because of its
importance, we have  investigated the properties of the lightest neutralino
in different patterns of supersymmetry breaking in the gaugino sector.
We have calculated lower limits on the masses of the neutralinos and
the charginos, taking into account the current experimental limits 
on the mass of the lightest  chargino.
Although these limits depend, through radiative corrections,
on parameters other than gaugino sector parameters, 
we have found that this  dependence is mild,
and thus the limits for the neutralino and chargino masses
can be considered to be relatively robust.


We have calculated an upper bound on the mass of the lightest neutralino 
as a function of the  lightest chargino mass.
We see that for the models of supersymmetry breaking considered 
in this paper, only in the mirage mediation model
with large $\alpha$, the upper bound found from the lower right hand
two-by-two part of the mass matrix becomes relevant.


The sum rule for the neutralino and chargino squared masses is one of the
distinguishing features of the supersymmetry breaking mechanism
in the gaugino sector.
For mSUGRA and AMSB the sign of the average squared mass difference
for charginos and neutralinos is different for $M_1$ larger than 
${\cal{O}}$(100 GeV), and for mirage 
mediation the parameter $\alpha$ can be deduced from the sum rule.


We have also discussed in detail the decay patterns of the neutralinos and
charginos in different models. An interesting result of our work is
that detection of neutralino and
chargino decay patterns gives important information
on the nature  of the underlying supersymmetry breaking mechanism, and may
help in identifying the correct supersymmetry breaking  pattern.


In Section V it was shown that the second lightest neutralino and the
lighter chargino are produced in large amounts in squark decays.
This is interesting, since a promising signal to detect weakly
interacting particles at Tevatron and at LHC is considered to be
the associated production $\tilde\chi^\pm_1\tilde\chi_2^0$,
see {\it e.g.} \cite{Barger:1998hp,Li:2007ih} and references therein.
Let us consider produced $\tilde\chi^\pm_1$, $\tilde\chi_2^0$ in view of
the cascade decays in
Figs.~(\ref{fig:brN2})-(\ref{fig:brC1}).
It is seen that in the studied breaking
patterns the largeness of the trilepton signal varies significantly.
In the mSUGRA pattern, $\tilde t_1$ decays to all the heavier 
neutralinos and charginos with nonnegligible branching fractions.
The contribution 
$\tilde t_1\rightarrow  \tilde\chi_1^+ b/\tilde\chi_2^0 t$ is at a few
percent level, but more events come from the decays of
$\tilde\chi_{3,4}^0$, $\tilde\chi_2^+$.
Thus from $\tilde t_1\bar{\tilde t}_1$ production there is an
additional contribution to the trilepton signal, accompanied by
a number of jets.
In the AMSB pattern, the enhancement of trileptons is significant.
$\tilde t_1$'s decay 60\% of the time to $\chi_2^0 t$ and 20 \% of
the time to $\chi_1^+ b$.
As soon as kinematically possible, the $\chi_2^0$ decays to a slepton and
lepton,  and $\chi_1^+$ decays leptonically 25\% of the time.
In mirage pattern, stops tend to decay directly to the lightest 
neutralino and no enhancement is expected.

Since the lightest neutralino is a possible candidate for the particle
dark matter, we have calculated its relic density in different
supersymmetry breaking models combining the information coming from decay
patterns.  While in the mSUGRA model typically a narrow range with the
observed relic density occurs, in the AMSB model the relic density
remains below the WMAP limit for the sub-TeV scale spectrum.  In
mirage mediation models the observed dark matter range is narrow and
close to the stop LSP region, unless the heavy Higgs resonance can be found.
We note that it is not necessary that neutralino is the only dark
matter particle, even if it were the lightest supersymmetric particle.
Furthermore, it is possible that the R-parity is broken at least
slightly in nature.  This would lead to the neutralino decay, even if
the breaking were so tiny that it would not show up in the
experiments.


\section{Acknowledgments}
KH and PT acknowledge the support by the Academy of Finland (Project No. 115
032).
The work of
J.L.~is supported in part by the Foundation for Fundamental 
Research of Matter (FOM) and the Bundesministerium f\"ur Bildung und
Forschung, Berlin-Bonn.
The work of P.N.P. is supported by the J. C. Bose National Fellowship,
the Board of Research in Nuclear
Sciences, and by the Council of Scientific and Industrial Research,
India.  He would like to thank the Helsinki Institute of Physics,
where part of this work was done, for its hospitality. 
PT thanks Magnus Ehrnrooth Foundation for support.


\appendix
\section{Chargino and Neutralino Mass Matrices}
In the wino-Higgsino basis
\bea
\psi_j^+ & = & (-i \lambda^+, \psi^1_{H_2}), \, \, \,  \, \,
\psi_j^-  =  (-i \lambda^-, \psi^2_{H_1}), \, \, \, \, \, j = 1, \,  2,
\label{chargestates}
\eea
where $\lambda^{\pm} = (1/\sqrt 2)( \lambda^1 \mp  \lambda^2)$,
and the superscripts  $1, \,  2$ refer to $SU(2)_L$ indices,
the chargino mass matrix can be written as~\cite{nilles}
\bea
{\mathcal
M_\pm} = \left(\begin{array}{cc} M_2 & {\sqrt 2} M_W \sin\beta\\
{\sqrt 2} M_W \cos\beta & \mu \\
\end{array} \right), \label{chargematrix}
\eea
where $M_2$ is the supersymmetry breaking
$SU(2)_L$ gaugino mass,  $\mu$ is the Higgs(ino)
mixing parameter, and $\tan\beta$ is the ratio of the vacuum expectation
values of the neutral components of the two Higgs doublets $H_2$ and
$H_1$.
We shall denote the eigenstates of the chargino mass
matrix~(\ref{chargematrix})   as
$\tilde \chi^{\pm}_1$ and $\tilde \chi^{\pm}_2$,  with
eigenvalues $M_{\tilde \chi^{\pm}_{i = 1, 2}},$ respectively.
The eigenvalues are most easily obtained from the diagonalization
of $ {\mathcal M_\pm}^\dagger {\mathcal M_\pm}$ resulting in the
squares of the chargino masses
\bea
M^2_{\chi^\pm_{1,2}}
  &=&\frac{1}{2}\left[M^2_2+\mu^2+2m^2_W
    \mp\sqrt{(M^2_2+\mu^2+2m^2_W)^2-4(M_2\mu-m^2_W\sin
2\beta)^2}\right].
\label{chmass}
\eea
On the other, in the bino-wino-Higgsino basis
\bea \psi^0_j = (-i\lambda',~ -i\lambda^3,~
\psi^1_{H_1},~ \psi^2_{H_2}),~~~ j = 1,~2,~3,~4,
\label{neut1}
\eea
where $\lambda'$ and $\lambda^3$ are the two-component gaugino
states corresponding to the $U(1)_Y$ and the third component of the
$SU(2)_L$ gauge groups, respectively, and  $\psi^1_{H_1}, \psi^2_{H_2}$
are the two-component Higgsino states,
the neutralino mass matrix can be written as~\cite{nilles}
\bea {\mathcal M_0}=\left(
\begin{array}{cccc} M_1 & 0 & -M_Z\cos\beta\sin\theta_W &
M_Z\sin\beta\sin\theta_W\\
0 & M_2 & M_Z\cos\beta\cos\theta_W & -M_Z\sin\beta\cos\theta_W\\
-M_Z\cos\beta\sin\theta_W& M_Z\cos\beta\cos\theta_W & 0 &-\mu\\
M_Z\sin\beta\sin\theta_W & -M_Z\sin\beta\cos\theta_W & -\mu & 0\\
\end{array} \right). \label{neutmatrix}
\eea
$M_1$ is the supersymmetry breaking $U(1)_Y$ gaugino mass, and
$g'$ and $g$ are the gauge couplings associated with the
$U(1)_Y$ and the $SU(2)_L$ gauge groups, respectively, with
$\tan\theta_W = g'/g$, and $M_Z^2 = (g^2 +g'^2)(v_1^2 + v_2^2)/2.$ The
neutralino mass matrix can be diagonalized by a unitary transformation
$N$ \bea N^{\dagger} {\mathcal M_0} N & = & {\mathcal M}^{\rm
  diagonal}_0.  \eea Assuming CP conservation, this transformation is
an orthogonal transformation.  We shall denote the eigenstates of the
neutralino mass matrix by $ \tilde \chi^0_1, \tilde \chi^0_2, \tilde
\chi^0_3, \tilde \chi^0_4$ with eigenvalues $M_{\tilde \chi^{0}_{i =
    1, 2, 3, 4}}$, labeled in order of increasing mass.  Explicit
expressions for these can be obtained, but these are not very
illuminating. 
The neutralinos are mixtures of gauginos and
Higgsinos 
\bea 
\label{neutmix}
\tilde \chi^0_i & = & N_{i1} \lambda' + N_{i2}
\lambda^3 + N_{i3} \psi^1_{H_1} + N_{i4} \psi^2_{H_2}.  \eea
\noindent
One can obtain  information on the neutralino masses by studying the
expansion of the neutralino mass matrix (\ref{neutmatrix})
in terms of  $M_Z/\mu$ for $M_Z \ll \mu$.
This  expansion is  obtained  most conveniently  by using the  basis
$(-i \tilde\gamma, -i \tilde Z^0, \tilde H^0_a,\tilde H^0_b)$, where
\bea
\tilde\gamma & = & \frac{1}{\sqrt{g^2 + g'^2}}(g' \lambda^3 + g
\lambda'),
\label{photino}\\
\tilde Z^0 & = & \frac{1}{\sqrt{g^2 + g'^2}}(g \lambda^3 - g' \lambda'),
\label{zino} \\
\tilde H^0_a  & = & \frac{1}{\sqrt{v_1^2 + v_2^2}}
(v_1 \psi_{H_1}^1 - v_2\psi_{H_2}^2) \label{higgsino1}, \\
\tilde H^0_b  & = & \frac{1}{\sqrt{v_1^2 + v_2^2}}
(v_2 \psi_{H_1}^1 + v_1\psi_{H_2}^2), \label{higgsino2}
\eea
are the photino, zino, and linear combinations of Higgsino states.  In
this
basis, after a similarity transformation (see {\it e.g.}
\cite{Huitu:2003ci}), the neutralino mass  matrix can
be written as
\begin{equation}
  \label{eq:diagneut}
\widetilde {\mathcal M_0} =  \left(
    \begin{array}{llll}
 {M_1} & 0 & -M_Z \cos \left(\beta -\frac{\pi }{4}\right) s_W  & M_Z
   \sin \left(\beta -\frac{\pi }{4}\right) s_W  \\
 0 & {M_2} & M_Z \cos \left(\beta -\frac{\pi }{4}\right) c_W  & -M_Z
   \sin \left(\beta -\frac{\pi }{4}\right) c_W  \\
 -M_Z \cos \left(\beta -\frac{\pi }{4}\right) s_W  & M_Z \cos
\left(\beta -\frac{\pi
   }{4}\right) c_W  & \mu  & 0 \\
 M_Z \sin \left(\beta -\frac{\pi }{4}\right) s_W  & -M_Z \sin
\left(\beta
   -\frac{\pi }{4}\right)c_W   & 0 & -\mu
\end{array}
\right).
\end{equation}
The mass matrix (\ref{eq:diagneut}) can  be diagonalized by using
perturbation theory for values of  $M_Z \ll \mu $.
For the case $ M_1 < M_2 $, which is what one  obtains in gravity
mediated
supersymmetry breaking~(see below), the mass of the
lightest neutralino can be written as, up to terms of $\mathcal O (M_Z/\mu)^2$,
\bea
m_{\chi^0_1} & = & M_1 - \frac{M_Z^2 s^2_W}\mu \sin 2\beta -
\frac {1} {\mu^2}
\left(M_Z^2 s^2_w M_1 +\frac{M_Z^4 s^2_W c^2_W}{M_2 -M_1}
\sin ^22\beta \right). \label{lightestN} \eea
Similarly, for
the second lightest neutralino ${\chi^0_2}$ one obtains
\bea
m_{\chi^0_{2}} & =  & M_2 - \frac{M_Z^2 c^2_W}\mu \sin 2\beta
- \frac {1} {\mu^2} \left(M_Z^2
c^2_W M_2 +\frac{M_Z^4 s^2_W c^2_W}{M_1-M_2 }\sin ^2 2\beta
\right), \label{2lightestN} \eea
where $ c^2_W \equiv \cos^2 \theta_W$ and $ s^2_W \equiv \sin^2
\theta_W$.
If instead we have  $ M_2 < M_1 $, a situation that arises in
anomaly mediated supersymmetry breaking models, Eq.~(\ref{2lightestN})
would represent the mass of the lightest
neutralino $\chi^0_{1}$, and Eq.~(\ref{lightestN}) would give the
formula
for the mass of the second lightest neutralino.  The dependence of the
lightest neutralino mass on the specific SUSY breaking scenario is due
to the fact that the ordering of the gaugino mass parameters is model
dependent.
The leading term in the remaining two neutralino masses is the
Higgs(ino) mixing parameter $|\mu|$.
Thus if $|\mu|$ value is small compared to $M_{1,2}$, Higgsino can
form a large or even dominant component of the
lightest neutralino, as can be seen from the mass formulae for the
remaining two neutralinos:
\bea
m_{\Neu 3} & = & \mu + \frac{M_Z^2}{2}(1+\sin 2\beta)  \frac {\mu -
  s^2_W M_2 - c^2_W M_1}{(\mu-M_1)(\mu-M_2)}
+ \frac {M_Z^4}{8 \mu^3} \cos^2 2 \beta,
\label{3lightestN}
\eea
and
\bea
m_{\Neu 4} & = & -\mu - \frac{M_Z^2}{2}(1-\sin 2\beta)  \frac {\mu +
  s^2_W M_2 + c^2_W M_1}{(\mu+M_1)(\mu+M_2)}
- \frac {M_Z^4}{8 \mu^3} \cos^2 2 \beta.
\label{4lightestN}
\eea

\end{document}